\documentclass[fleqn,usenatbib,useAMS]{mnras}

\usepackage{subfiles}
\usepackage{amsmath, graphicx, amssymb,xifthen,lipsum,comment,xcolor}
\newcommand{\deleted}[1]{}
\newcommand{\lowestM}{$\dot m\sim 10^{-6}$}
\newcommand{\highestM}{$\dot m\sim 10^{0}$}
\usepackage[T1]{fontenc}
\usepackage{ae,aecompl}

\usepackage{newtxtext,newtxmath}
\title[Thin Disc Inner Structure]{The inner structure and thermodynamics of thin accretion discs}

\author[Hankla, Dexter, \& Scepi]{Amelia M. Hankla,$^{1}$%
\thanks{E-mail: \href{mailto:lia.hankla@gmail.com}{lia.hankla@gmail.com}, \href{mailto:ahankla@umd.edu}{ahankla@umd.edu}}%
\thanks{NASA Hubble Einstein Fellow}
Jason Dexter$^{2,3}$,  
Nicolas Scepi$^4$
\\
$^{1}$Department of Astronomy, University of Maryland, 4296 Stadium Dr. Ste 1113, College Park, MD 20742, USA\\
$^{2}$JILA, University of Colorado and National Institute of Standards and Technology, 440 UCB, Boulder, CO 80309-0440, USA\\
${^3}$Department of Astrophysical and Planetary Sciences, University of Colorado, 391 UCB, Boulder, CO 80309, USA\\
${^4}$Univ. Grenoble Alpes, CNRS, IPAG, 38000 Grenoble, France}

\date{Last updated \today; in original form \today}

\pubyear{2025}

\begin{document}
\label{firstpage}
\pagerange{\pageref{firstpage}--\pageref{lastpage}}
\maketitle

% Abstract of the paper
\begin{abstract}
Using three-dimensional general relativistic magnetohydrodynamic simulations with electron and proton thermodynamics and an electron cooling function, we probe the inner radial and vertical structure of weakly magnetized geometrically thin accretion discs around rapidly spinning black holes. 
We find that the thin, cold disc transitions to a thick, hot accretion flow at a radius dependent on the mass accretion rate due to proton-electron Coulomb decoupling.
At high accretion rates, the disc truncates close to the innermost stable circular orbit $r\approx2r_g$, demonstrating that even in the canonical thin disc model, the plunging region should be treated with two-temperature physics.
At intermediate accretion rates, the transition radius moves outward by a factor of two to $r\approx 5r_g$, forming a radiatively inefficient inner flow.
The simulations also reveal extended cooling along the surface of the disc out to $\sim10r_g$, with 40\% of the total cooling at intermediate accretion rates occurring above the disc body.
Two-temperature effects also impact the emission from within the plunging region of the black hole, leading to less thermal emission than predicted by single-temperature models.
These results have implications for X-ray binary state transitions, the physical origin of the X-ray corona, and spin measurements that rely on determining the location of the innermost stable circular orbit. 
\end{abstract}

% Select between one and six entries from the list of approved keywords.
% Don't make up new ones.
\begin{keywords}
accretion -- black hole physics -- X-rays: binaries
\end{keywords}

%%%%%%%%%%%%%%%%%%%%%%%%%%%%%%%%%%%%%%%%%%%%%%%%%%

\section{Introduction} \label{sec:intro}
Geometrically thin, optically thick accretion discs \citep{ss73,novikovthorne} are generally thought to power luminous black hole X-ray binaries (XRBs) and active galactic nuclei (AGN). 
Thin disc models assume that the disc electrons efficiently radiate the viscously dissipated energy that heats both electrons and protons.
The protons in thin disc models cool due to efficient Coulomb collisions with electrons enabled by high densities. 
Standard models often assume that the flow truncates abruptly at the innermost stable circular orbit (ISCO), with no dissipation or radiation inside of that radius. 
However, observations~\citep{fabian2020}, analytic models~\citep{gammie1999, agol2000}, and numerical simulations~\citep{noble2009,zhu2012,avara2016, scepi2024, dhang2025} have suggested that magnetic torques may result in dissipation and non-negligible amounts of radiation coming from within the plunging region.
These models have implicitly assumed that the electrons and protons remain well-coupled within the ISCO, despite hints to the contrary~\citep{zhu2012}. 
Analytic estimates showed that electrons and protons may decouple within the ISCO, potentially explaining the $>10$ keV tail observed in the soft state of XRBs~\citep{hankla2022}.  

Numerical simulations have attempted to realize global models of a thin accretion disc using general relativistic ideal magnetohydrodynamics (GRMHD). 
Since the disc cools by radiation, such simulations have either used approximate forms of radiative transfer~\citep{liska2022} or idealized the cooling as a source term that removes thermal energy from the gas~\citep{shafee2008,noble2009,penna2010}.
These prescriptions often assume perfect coupling between electrons and ions by treating the gas as a single fluid.
In some cases, radiation is then postprocessed onto the disc to calculate predictions for XRB spectra~\citep{kinch2020}.
Two-temperature thin discs that evolve electrons as a separate fluid~\citep{ressler2015} have been mainly studied in the context of strong magnetic fields at high accretion rates, where magnetic torques cause the thin disc to transition to a hot flow at small radii~\citep{liska2022}.
These magnetically-arrested (MAD) discs can appear observationally truncated at a radius not necessarily related to the radius at which the disc becomes MAD~\citep{scepi2024b}.%

Proton-electron coupling through Coulomb collisions plays an important role in accretion-rate-dependent models for XRB state transitions. 
In the ``disc truncation'' model, the thin accretion disc transitions to a hot accretion flow at radii $\gg r_{\rm ISCO}$ at intermediate accretion rates~\citep{esin1997, done2007}.
Whereas the soft state of XRBs is well described by a cold, thin disc, the hard state likely originates from photons Compton scattering off a geometrically thick, optically-thin accretion flow~\citep{narayan1994, yuan2014}. 
Geometrically thick, optically-thin accretion flows have low densities, inefficient Coulomb coupling, and therefore very little proton cooling.
In these discs, the electron cooling does not influence the disc structure, motivating the evolution of a single-temperature flow whose proton-to-electron temperature ratio and thus radiation is calculated during postprocessing~\citep{moscibrodzka2009}.
GRMHD simulations that treat radiation through Monte Carlo methods have previously suggested that electron dynamics become important at accretion rates higher than expected by analytic models~\citep{ryan2017}, potentially collapsing the hot flow to a thin disc~\citep{dexter2021}.

Because of the difficulty in treating both radiation and thermodynamics, the transition from thick to thin disc has only been studied in local context or with a pre-determined truncation radius.
Radiation becomes difficult to model in these intermediate accretion rate regimes.
Monte Carlo methods become computationally intractable as the photon optical depth increases the number of scatterings, whereas treating radiation as a fluid in optically-thin regions may not accurately treat anisotropic and non-local Compton cooling.
Therefore, numerical simulations have attempted to access this radial transition by putting in the transition by hand and examining the resulting structure~\citep{hogg2017,hogg2018}.
A complementary approach is to examine the vertical structure alone using local shearing box simulations.
Although these simulations found vertical decoupling of protons and electrons~\citep{bambic2024}, by their nature they cannot probe a radial transition in the accretion disc structure.

In this work, we allow Coulomb collisions to determine the disc vertical and radial structure by treating radiation as an electron-only cooling function.
By removing energy from the electron fluid rather than the total gas, protons cool exclusively through Coulomb energy exchange.
This electron cooling function therefore forms a thin disc at high densities and a thick disc at low densities, allowing for the possibility of disc truncation.
We describe the implementation of the electron cooling function and implicit treatment of Coulomb collisions in Sec.~\ref{sec:methods}.
We then examine the impact of two-temperature thermodynamics on a canonical thin accretion disc, showing that even at high accretion rates, the plunging region decouples into a two-temperature flow (Sec.~\ref{ssec:highM-ISCO}).
Next, we introduce a suite of simulations that probe the accretion disc's radial structure as a function of accretion rate, finding that the disc transitions to a thick flow at $r\gtrsim r_{\rm ISCO}$ for intermediate accretion rates (Sec.~\ref{ssec:parameterscan}).  
Finally, we discuss implications for observing emission from the plunging region in canonical thin discs, XRB state transitions, and the nature of the X-ray corona in Sec.~\ref{sec:discussion}.

\section{Methods} \label{sec:methods}
The simulations in this work use $\texttt{nubhlight}$~\citep{miller2019}, an extension of the public GRMHD code $\texttt{ebhlight}$~\citep{ryan2015,ryan2017}, which solves the general relativistic ideal magnetohydrodynamic equations using the~$\texttt{HARM}$ scheme~\citep{gammie2003,noble2006}. 
An electron fluid, evolved separately from the background fluid~\citep{ressler2015}, heats via Coulomb collisions and grid-scale dissipation. 
The viscously dissipated energy partitions into electron and proton heating according to particle-in-cell simulations of 2D magnetic reconnection with a moderate guide field; see~\citet{werner2018} Eq. 3.
This prescription gives a fraction $\delta_e$ between 1/4 and 1/2 of the dissipated viscous energy to electrons according to the magnetization of the plasma $\sigma_i$ in a given simulation cell: $\delta_e=0.25(1+\sqrt{\sigma_i/5/(2+\sigma_i/5)}$.
The cooling function described below affects only this electron fluid, in contrast to single-fluid GRMHD cooling functions. 

\subsection{Implementation of the electron cooling function} \label{ssec:ecooling}
Both $\texttt{nubhlight}$ and $\texttt{ebhlight}$ include radiative transfer with an explicit Monte Carlo method. 
We do not use this functionality, and instead set the radiation force $G_\mu$ directly. 
This method assumes that all radiative cooling is optically thin, leaving the system without interacting with the gas. 
Assuming isotropic radiation in the fluid's rest frame, the cooling function removes energy from the local energy conservation equation following~\citet{noble2009} via:
\begin{equation}
    G_\mu=-Q_{\rm cool}u_\mu,
\end{equation}
where the radiation force $G_\mu=\nabla_\mu T^\mu_\nu$, $u_\mu$ is the fluid four-velocity. 
The quantity $Q_{\rm cool}$ is the energy per volume radiated per unit time in the fluid's frame. 
We take
\begin{equation}
    Q_{\rm cool}=\frac{2u_e\left(T_e/T_{e, {\rm target}}\right)^{0.5}}{t_{\rm cool}},
\end{equation}
where $u_e$ is the electron energy density. 
The cooling function shuts off when $T_e < T_{e, {\rm target}}$ by setting $Q_{\rm cool}=0$. 
To avoid cooling the jet, the cooling function also sets $Q_{\rm cool}=0$ when the plasma magnetization $\sigma \equiv B^2/\rho > 1$. 

We assume electron cooling is always efficient, e.g., that the cooling time is short.
We set the electron cooling time to the shortest timescale in the system: $t_{\rm cool}=\Omega^{-1}$, where $\Omega=(r^{3/2}+a)^{-1}$ is the Keplerian orbital frequency. 
Inside the ISCO, $\Omega$ becomes the orbital frequency of a particle with the same angular momentum and specific energy as a circular orbit at the ISCO~\citep[Eq. 16]{noble2009}.
To mimic the impact of inverse Compton cooling in an optically-thin flow, we set the target temperature to a constant value of $10^{10}$ K. 
Although the electrons would more realistically cool down until they reach the non-relativistic regime at $T_e\sim10^9~{\rm K}$, such small gas temperatures would lead to a disc with scale height ratio $H/r\sim0.02$, which is beyond this paper's numerical capabilities to resolve.
We ensure that supercooling of the gas remains rare, occurring in less than $0.01\%$ of the simulation cells.

\subsection{Implementation of implicit Coulomb energy exchange}
In the low density regime that \texttt{nubhlight} was originally intended for, Coulomb collisions can be treated explicitly.
However, because we aim to extend \texttt{nubhlight} to higher accretion rates and thus higher densities, we must treat Coulomb collisions implicitly.
In each call of the Coulomb heating function, the code solves the equation:
\begin{equation}
    \frac{{\rm d}u_e}{{\rm d}t}=\frac{Q_{\rm coul}(u_e,u_g-u_e)}{u^0},
\end{equation}
where $Q_{\rm coul}$ is the energy exchange rate between protons and electrons given by~\citet{stepney1983} as a function of $u_e$ and $u_g$: the electron and total gas internal energy density, respectively.
We solve this equation implicitly by discretizing it via the Crank-Nicolson method and solving it with Brent's method.
Brent's method requires an interval over which the right-hand side switches sign. 
With this problem, the choice of interval is simple: assuming $T_e<T_p$ initially, the minimum solution is $T_e$ and the maximum is the equilibrium temperature $T_0$ where $T_e=T_p$.
The implementation sets values of $u_e$, $u_g$, and $Q_{\rm coul}$ after viscous heating has already been applied to the total gas and electron fluid.
To speed up computation, the implementation enters the single-temperature regime when the fractional change in $u_e$ is larger than 10 and $T_e$ and $T_p$ are both within $0.1\%$ of $T_0$.
Instead of implicitly solving for $T_e$ and $T_p$, both temperatures are set to the equilibrium temperature $T_0$. 
The implementation switches between an implicit and explicit evolution depending on whether the fractional change in $u_e$ is greater or less than $10^{-5}$.
See Appendix~\ref{app:tt-tests} for implementation tests.

\subsection{GRMHD Simulations}\label{ssec:grmhd}
The GRMHD simulations start with a gas torus in hydrostatic equilibrium~\citep{fishbonemoncrief} and accrete due to the MRI introduced by seeding the torus with a weak magnetic field.
The gas torus has an inner radius of $20r_g$, with the pressure maximum located at $41r_g$.
For all simulations, the black hole has a mass of $10M_\odot$ and a dimensionless spin $a=0.9375$.
The initial magnetic field has a poloidal configuration with a maximum $\beta=100$.

We label simulations by their approximate accretion rates normalized to the Eddington accretion rate $\dot m\equiv \dot M/\dot M_{\rm Edd}=\dot M/(L_{\rm Edd}/0.1c^2)$, as scaled from the code's mass unit: M6 for $\dot m\sim 10^{-6}\dot m_{\rm Edd}$, M2 for $\dot m\sim 10^{-2}\dot m_{\rm Edd}$, and so on.
The lowest accretion rate (M6) provides a test case in that Coulomb collisions are subdominant to viscous heating everywhere in the disc body.
All other simulations restart from M6's restart file at $2\times10^4~r_g/c$, with the torus mass (and hence density and magnetic field strength) rescaled to a higher value in order to produce a larger value of $\dot{m}$. 
Because of the long timescales involved in coming to equilibrium through Coulomb collisions, simulations other than M6 are run with Coulomb collisions enhanced by a factor of 10 for the period $2\times10^4<tc/r_g<3\times10^4$; see Appendix~\ref{app:enhancedQ}.
From $3\times10^4~r_g/c$ onwards, the simulations use the correct, unenhanced Coulomb collision rates.

The simulations have $[320, 256, 160]$  cells in the $[r,\theta,\phi]$ directions.
The simulation domain extends $2\pi$ radians in $\pi$, $\pi$ radians in $\theta$, and from within the event horizon out to $r=1000r_g$.
Unless otherwise noted, the simulation data are time-averaged over the last $2000r_g/c$, with the final times given in Table~\ref{tab:last-times}.
The simulations use the standard drift-frame floors: density and total internal energy density are added in the drift frame when their values dip beneath the floor values $\rho_{\rm floor}={\rm max}[b^\mu b_\mu/50, 10^{-6}(r/r_g)^{-2}]$ and $u_{\rm floor} = {\rm max}[b_\mu b^\mu/2500,~10^{-8}(r/r_g)^{-2\gamma}]$, respectively~\citep{ressler2015, ressler2017}.
Throughout, we use code units unless otherwise specified.

\begin{table}
\centering
\begin{tabular}{l||c|c|c|c}
 & $(t_{\rm final}c/r_g)/10^3$ & $\dot M/\dot M_{\rm Edd}$ &  $Q_\theta$ & $Q_\phi$\\ \hline\hline
M6 (\lowestM) & 42.4 & $9.7\times10^{-7}$ & 25 & 25\\ \hline
M2 ($\dot m\sim 10^{-2})$ & 48.4 & $1.5\times10^{-2}$ & 8 & 14\\ \hline
M1 ($\dot m\sim 10^{-1})$  & 39.9 & $1.2\times10^{-1}$ & 8 & 13\\ \hline
M0 (\highestM) & 38.2 & $1.3\times10^0$ & 7 & 12\\ \hline
\end{tabular} 
\caption{Final simulation times for each high-resolution run. Corresponding accretion rate $\dot M$ (Eq.~\ref{eq:mdot}) normalized to the Eddington accretion rate $\dot M_{\rm Edd}=L_{\rm Edd}/(0.1c^2)$ and MRI quality factors $Q_\theta$ and $Q_\phi$ are time-averaged over the last $2000r_g/c$. Quality factors are also averaged over $r_{\rm EH}<r<20r_g$.}\label{tab:last-times}
\end{table}

\subsection{Diagnostics}
This section outlines several diagnostic measurements used throughout this paper.
Angle brackets denote density-weighted shell-averages:
\begin{equation}
    \langle f\rangle=\frac{\int\int_{\sigma<1}~f\rho~\sqrt{-g}d\theta d\phi}{\int\int_{\sigma<1}~\rho~\sqrt{-g}d\theta d\phi},\label{eq:shell-avg-def}
\end{equation}
whereas double angle brackets indicate volume-integrated quantities:
\begin{equation}
    \langle\langle f\rangle\rangle = \iiint_{\sigma<1} f(r,\theta,\phi)\sqrt{-g}~{\rm d}\theta~{\rm d}\phi~{\rm d}r.
\end{equation}
To exclude the floor-dominated region within the jet, the integration is performed over regions with magnetization $\sigma\equiv b_\mu b^\mu/\rho<1$.

The mass accretion rate is defined as the mass flux at the event horizon and we refer to its value time-averaged over the last $2\times10^3r_g/c$, as outlined in Table~\ref{tab:last-times}:
\begin{equation}
    \dot M\equiv \dot M(r_{\rm EH})=-\iint \rho u^r\sqrt{-g}~{\rm d}\theta~{\rm d}\phi. \label{eq:mdot}
\end{equation}

The disc scale height $H$ is defined as the variance of the density moment~\citep{noble2009}:
\begin{equation}
    H \equiv\left(\langle z^2\rangle -\langle z\rangle^2\right)^{1/2} \label{eq:h-over-r}
\end{equation}
where $z=r\cos\theta$ and both $\langle z^2\rangle$ and $\langle z\rangle$ are defined using Eq.~\ref{eq:shell-avg-def}.
Using the variance rather than the mean better accounts for times when the disc is not symmetric about the midplane.

The midplane optical depth is defined as integrating the line-of-sight along constant radius, going from the pole to the midplane and averaging the two hemispheres:
\begin{align}
    %\tau_{es}(r,\theta,\phi)=\int_{0}^{\theta} \kappa\rho(r,\theta',\phi)~\sqrt{g_{\theta\theta}}{\rm d}\theta' & \text{if}~0<\theta<\pi/2\\
    \tau_{es}(r)=\frac12\int_0^{2\pi}~{\rm d}\phi\left( \int_{0}^{\pi/2}\kappa_{\rm es}\rho(r,\theta,\phi)~\sqrt{g_{\theta\theta}}{\rm d}\theta\right.\\
     \left.+\int_{\pi}^{\pi/2}\kappa_{\rm es}\rho(r,\theta,\phi)~\sqrt{g_{\theta\theta}}{\rm d}\theta\right)
\label{eq:opt-depth}
\end{align}
Here, $\kappa_{\rm es}=0.4~{\rm cm^2/g}$ is the Thomson opacity.
In this equation, we include only the optical depth due to Thomson scattering.
The effective optical depth that includes both scattering and absorption is likely up to a factor of 10 different close to a stellar-mass black hole.

We define various timescales within the system to determine their relative importance in the accretion discs' evolution as a function of radius.
We define the infall time $t_{\rm infall}$ as the time for gas at a given radius to fall into the black hole.
The other proton and electron timescales take the ratio of the species' internal energy density $u_i$ to the energy density exchange rate $Q$, where $i\in[e,~p]$ for electrons or protons.
The physical processes involved for these timescales are electron cooling $t_{\rm cool}^e$, proton/electron Coulomb equilibration $t_{\rm Coul}^{i}$, and proton/electron heating due to viscous dissipation $t_{\rm heat}^i$.
Explicitly, the timescales are defined as:
\begin{align}
    t_{\rm infall}(r)&=\int_{r_{\rm EH}}^r~\frac{1}{u^r(r')}~{\rm d}r'& 
        t_{\rm cool}^e&=u_e/Q_{\rm cool}\label{eq:tinfall}\\
    t_{\rm Coul}^p&=u_p/Q_{\rm coul}&
        t_{\rm Coul}^e&=u_e/Q_{\rm Coul} \label{eq:tCoul}\\
    t_{\rm visc}^p&=u_p/Q_{\rm visc}^p&
        t_{\rm visc}^e&=u_e/Q_{\rm visc}^e. \label{eq:tVisc}   
\end{align}
The relative ordering of these timescales will determine the accretion disc's structure.

Throughout, we will define observed ``luminosities'' interior to a given radius as integrations of energy dissipation rates.
The canonical luminosity integrates over the energy $Q_{\rm cool}$ removed due to radiation:
\begin{equation}
    L(<r)=\int_{\phi=0}^{2\pi}\int_{\theta=0}^\pi\int_{r_{\rm EH}}^r -Q_{\rm cool}(r',\theta,\phi)u_t(r',
    \theta,\phi)\sqrt{-g}~{\rm d}r'~{\rm d}\theta~{\rm d}\phi. \label{eq:integrated-L}
\end{equation}
We also define $L_{\rm visc}(<r)$ the same as Eq.~\ref{eq:integrated-L} but with $Q_{\rm visc}$ rather than $Q_{\rm cool}$. 
As a shorthand, we define $L(<r\to\infty)\equiv L_\infty$.
Because these luminosities are not ray-traced, they do not include effects such as Doppler boosting.

For comparison to the Novikov-Thorne prediction, we will often refer to the radial derivative of the luminosity normalized to the mass accretion rate, i.e.
\begin{align}
    \mathcal{L}(r)&\equiv\frac{dL(<r)/\dot M}{dr} \nonumber\\
    &=\frac{1}{\dot M}\int_{\phi=0}^{2\pi}\int_{\theta=0}^\pi -Q_{\rm cool}(r,\theta,\phi)u_t(r,
    \theta,\phi)\sqrt{-g}~{\rm d}\theta~{\rm d}\phi. \label{eq:dLdr}
\end{align}
Here, $\mathcal{L}(r)$ is the function $f(r)$ given in~\citet[Eq. 15n]{page1974}.

% -----------------------------------------------------
% -----------------------------------------------------
\begin{figure*}
    \centering
    \includegraphics[width=0.8\linewidth]{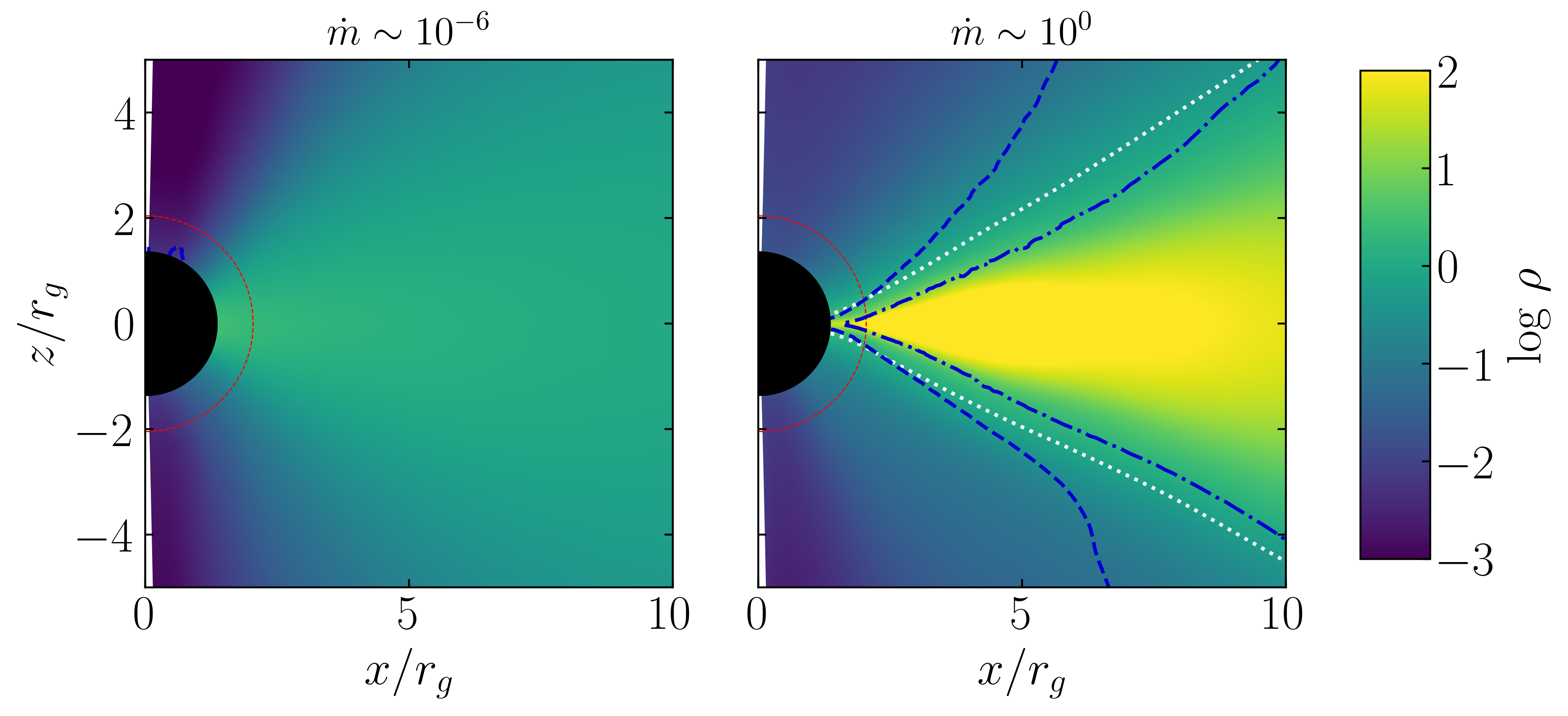}
    \caption{Azimuthally-averaged density structure of the low accretion rate simulation M6 (left) and highest accretion rate simulation M0 (right). The white dotted contour shows $\tau_{\rm es}=10$, while the dark blue dash-dot and dashed lines show $T_p/T_e=2$ and 10, respectively. The red dashed line shows the equatorial ISCO radius. Time-averaged over the last $2000r_g/c$.}    \label{fig:slices-case}
\end{figure*}
\section{Results}
\subsection{The Plunging Region of a Thin Disc} \label{ssec:highM-ISCO}
\begin{figure*}
    \centering
    \includegraphics[width=0.8\linewidth]{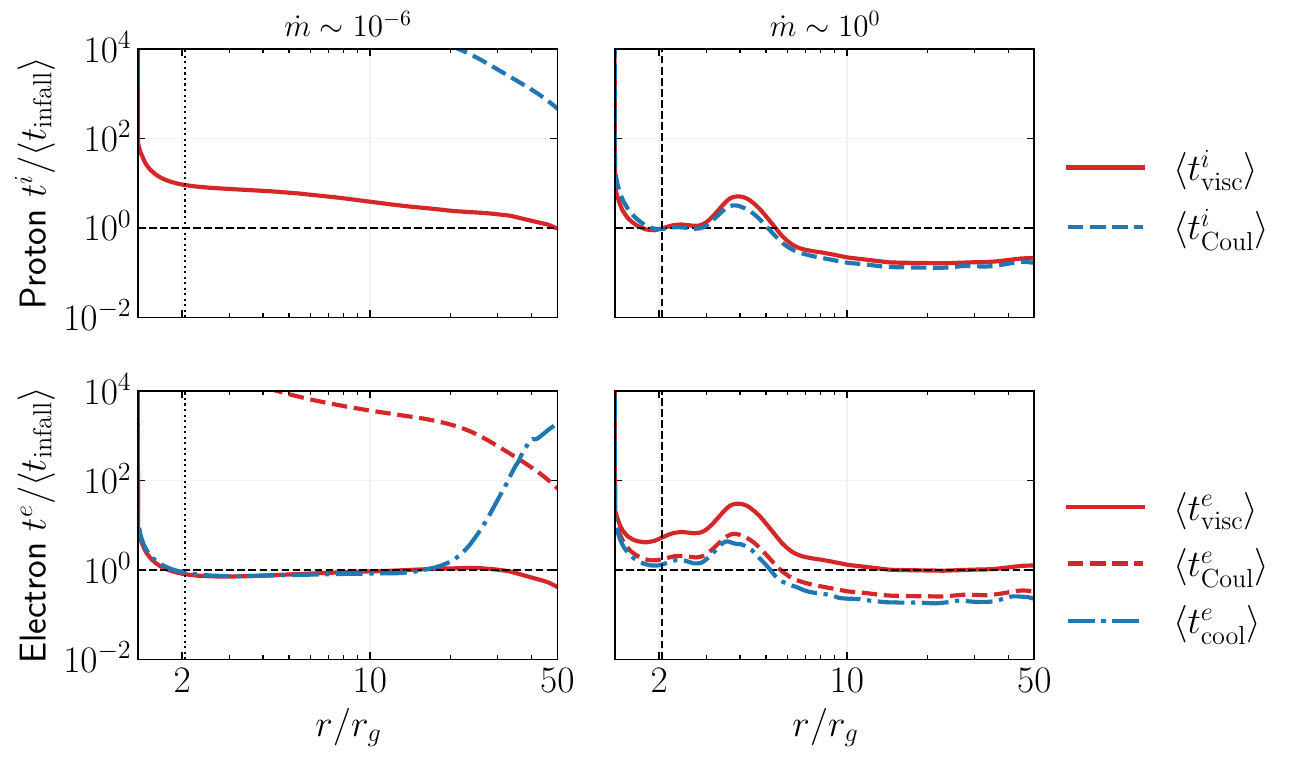}
    \caption{Demonstration of the proton and electron timescales relative to the infall time. The left column shows the low accretion rate simulation M6; the right column shows the canonical high accretion rate simulation M0. Rows show proton timescales (top) or electron timescales (bottom). In all panels, the dashed horizontal black line shows the infall time. Red indicates that the physical process heats the indicated species, while processes in blue cool the indicated species. Time-averaged over the last $2000r_g/c$. Vertical dotted line shows the location of the ISCO.}
    \label{fig:timescales-demo}
\end{figure*}
We begin with a case study comparing a low accretion rate simulation where Coulomb collisions are negligible (M6) to a high accretion rate simulation (M0) where Coulomb collisions collapse the initially thick torus into a thin disc with the target scale height. 
Fig.~\ref{fig:slices-case} shows the difference in density structure for these two simulations.

The timescales outlined in Eqns.~\ref{eq:tinfall}-\ref{eq:tVisc} demonstrate the differences between the thick and thin disc.
For the thick, hot accretion flow M6 (Fig.~\ref{fig:timescales-demo} left column), the Coulomb energy exchange rate for the protons far exceeds the infall time for $r<100r_g$, showing that Coulomb collisions are indeed irrelevant in determining the thick disc's inner structure at low accretion rates.
The proton heating timescale also stays above the infall time until $r\gtrsim50r_g$, consistent with the ADAF model's idea that the gas advects rapidly inward.
For electrons, the viscous heating and electron cooling rate precisely balance each other and the infall time for $r\lesssim20r_g$, showing that the electrons are in thermal equilibrium. 
The timescale for electron heating due to Coulomb collisions is more than four orders of magnitude greater than the infall time, showing that Coulomb collisions can be neglected for electrons at low accretion rates.

In contrast, the thin accretion disc's structure changes as Coulomb collisions become important in most regions of the disc (Fig.~\ref{fig:timescales-demo} right column). 
For large radii, the proton cooling timescale due to Coulomb collisions equals the proton heating timescale due to viscous dissipation, and both timescales happen efficiently before material falls onto the black hole: $t^i_{\rm visc}\simeq t^i_{\rm Coul}\ll t_{\rm infall}$.
Similarly, the electron heating and cooling timescales at large radii are also faster than the infall time. 

Unlike the thick disc, the thin accretion disc's structure changes as orbits within the ISCO shift from predominately azimuthal to predominately radial.
In this region, the infall time becomes the shortest timescale in the system regardless of accretion rate.
Both the proton viscous heating time and the proton Coulomb equilibration timescales become longer than the infall time, i.e.
\begin{equation}
    t_{\rm infall}<t_{\rm visc}^i<t_{\rm Coul}^i.~\label{eq:thin-timescales}
\end{equation}
This change in the timescale ordering has profound results for the structure of thin discs within the ISCO, as we now discuss.

\subsubsection{The Plunging Region Is Two-Temperature}
\begin{figure}
    \centering
    \includegraphics[width=0.95\linewidth]{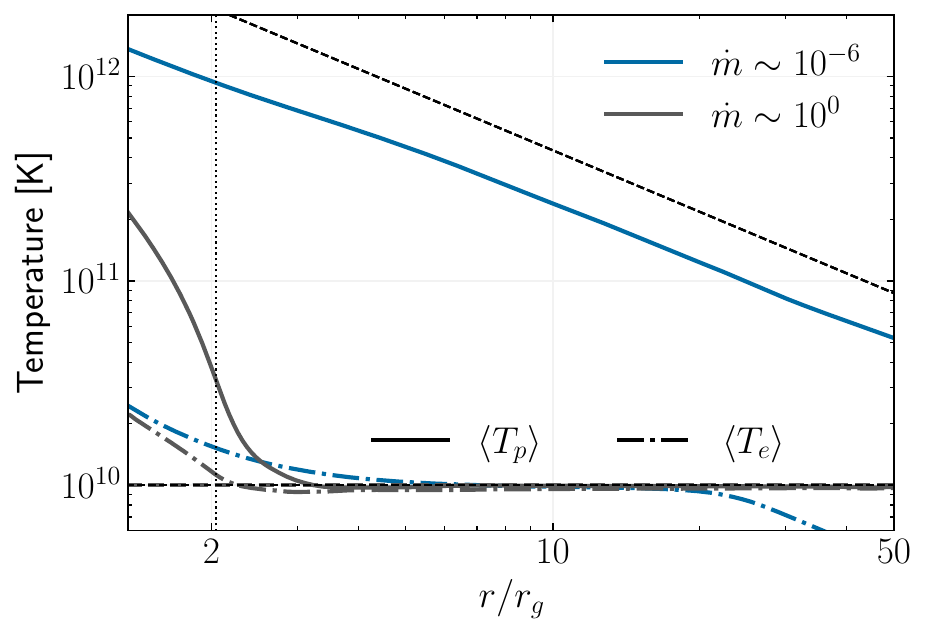}
    \caption{Comparison of electron and proton thermodynamics for low (M6) vs. high (M0) accretion rate. Coulomb collisions are insufficient to cool protons in M6 (blue solid line). At high accretion rate M0, the disc reaches the single temperature regime for $r\gtrsim4r_g$, but Coulomb collisions within this radius are insufficient to cool the protons down to the electron temperature. Time-averaged over the last $2000r_g/c$. Vertical dotted line shows the location of the ISCO.}
    \label{fig:temp-demo}
\end{figure}

Because the infall time becomes shorter than the Coulomb equilibration time, any difference in proton and electron temperatures will not disappear by the time the gas accretes.
Adiabatic heating will naturally lead to a temperature difference since the relativistic electrons have an adiabatic index of $\gamma_e=4/3$, whereas the non-relativistic protons have an adiabatic index $\gamma_p=5/3$.
Therefore, we expect the flow within the plunging to become two-temperature. 
Fig.~\ref{fig:temp-demo} demonstrates the two-temperature nature of both the hot accretion flow (\lowestM) and the high accretion rate, thin disc (\highestM).
The protons in the low accretion rate simulation are virial, with electrons at or close to the imposed target temperature.
In contrast, the protons in the high accretion rate simulation have the same temperature as the cooled electrons for $r\gtrsim r_{\rm ISCO}$.
At the ISCO, the proton temperature increases by a factor of 10 above the electron temperature because Coulomb collisions cannot equilibrate the two species before they accrete onto the black hole.
The protons reach temperatures of $2\times10^{11}~{\rm K}$ within $0.1r_g$ of the event horizon.
Viscous heating also contributes to the temperature difference between protons and electrons at high accretion rates because $t_{\rm visc}^i<t_{\rm Coul}^i$ (Fig~\ref{fig:timescales-demo} right).
Although viscous heating is slow compared to the infall time, it still contributes to the heating of both electrons and protons.

The shortened infall time also slightly affects the electron temperature within the ISCO because the cooling time becomes longer than the infall time (Sec.~\ref{ssec:ecooling}).
Fig.~\ref{fig:temp-demo} shows that $T_e$ rises about a factor of 2 above the target for both the low and high accretion rate simulations. 
For radii $r\gtrsim20r_g$, Coulomb collisions increase the electron temperature to the target temperature in the high accretion rate simulation.

\subsubsection{Flow Structure Within the Plunging Region}
\begin{figure}
    \centering
    \includegraphics[width=0.95\linewidth]{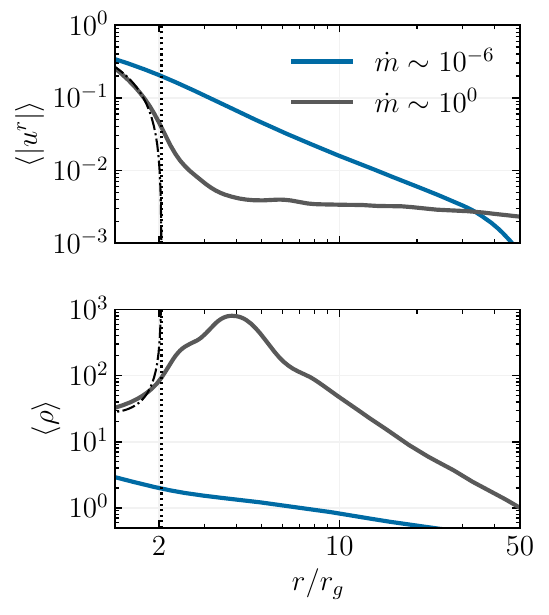}
    \caption{Comparison of flow velocity $u^r$ absolute magnitude (top) and rest-mass density $\rho$ (bottom) for low (M6; blue) vs. high (M0; gray) accretion rate. Time-averaged over the last $2000r_g/c$. Vertical dotted line shows the location of the ISCO. Dot-dash black lines show predictions from~\citet{gammie1999} with $F_{\theta\phi}=2$ and $a=0.9375$.}
    \label{fig:Gammie-demo}
\end{figure}

The thin disc structure inside the ISCO closely follows predictions for previous single-temperature work.
As shown in Fig.~\ref{fig:Gammie-demo}, the radial four-velocity $u^r$ remains small and relatively constant until $r\approx2r_{\rm ISCO}$, when it begins to rise rapidly.
At the same location, the mass density $\rho$ begins to drop rapidly.
Both quantities roughly follow single-temperature predictions~\citep{gammie1999}, with deviations as in previous single-temperature GRMHD simulations that are due to thermal effects yielding non-zero $u^r$ at the ISCO~\citep{penna2010}.
Because of the timescale ordering in Eq.~\ref{eq:thin-timescales}, the density scale height of the thin disc increases inside the ISCO (see Sec.~\ref{sssec:structure}).
In comparison, the thick disc shows no discernible change at the ISCO in either quantity.
As expected, the thick disc's radial velocity is generally higher than the thin disc's, whereas the density is lower than the thin disc's. 

% \subsubsection{Magnetic Field Structure in the Plunging Region}
% For low Mdot, Btheta and Bphi dominate within the ISCO, Bphi dominates outside the ISCO.
% For high Mdot, Bphi dominates both other components.
% Br increases by about an order of magnitude relative to the low mdot simulation.
% Bmag approximately the same for both simulation (no dragging in flux for larger radii; too short time)
% \begin{figure}
%     \centering
%     \includegraphics[width=0.95\linewidth]{shell_compare-B-components.pdf}
%     \caption{Comparison of magnetic field components for low (M6; left) vs. high (M0; right) accretion rate. Time-averaged over the last $2000r_g/c$. Vertical dotted line shows the location of the ISCO.}
%     \label{fig:B-demo}
% \end{figure}

% ------------------------------------------------------- %

\subsection{Dependence of Disc Properties on Accretion Rate} \label{ssec:parameterscan}
%Implementing an electron-only cooling function permits probing the intermediate mass accretion rate regime in a well-controlled manner.
We next investigate the impact of Coulomb collisions on the structure of the disc by holding the electron target temperature at a constant value of $10^{10}~{\rm K}$ and varying the scaling of the mass unit, thereby adjusting the accretion rate.
Because the Coulomb energy exchange rate depends quadratically on the mass unit, increasing the accretion rate will increase the importance of Coulomb collisions throughout the disc.
In this section, we explore where Coulomb collisions become important enough to affect the disc structure in the radial and vertical directions.
All accretion rates are relatively constant over time; the time-averaged values are listed in Table~\ref{tab:last-times}.

\subsubsection{Thermodynamics} \label{sssec:thermodynamics}
\begin{figure*}
    \centering
    \includegraphics[width=0.95\linewidth]{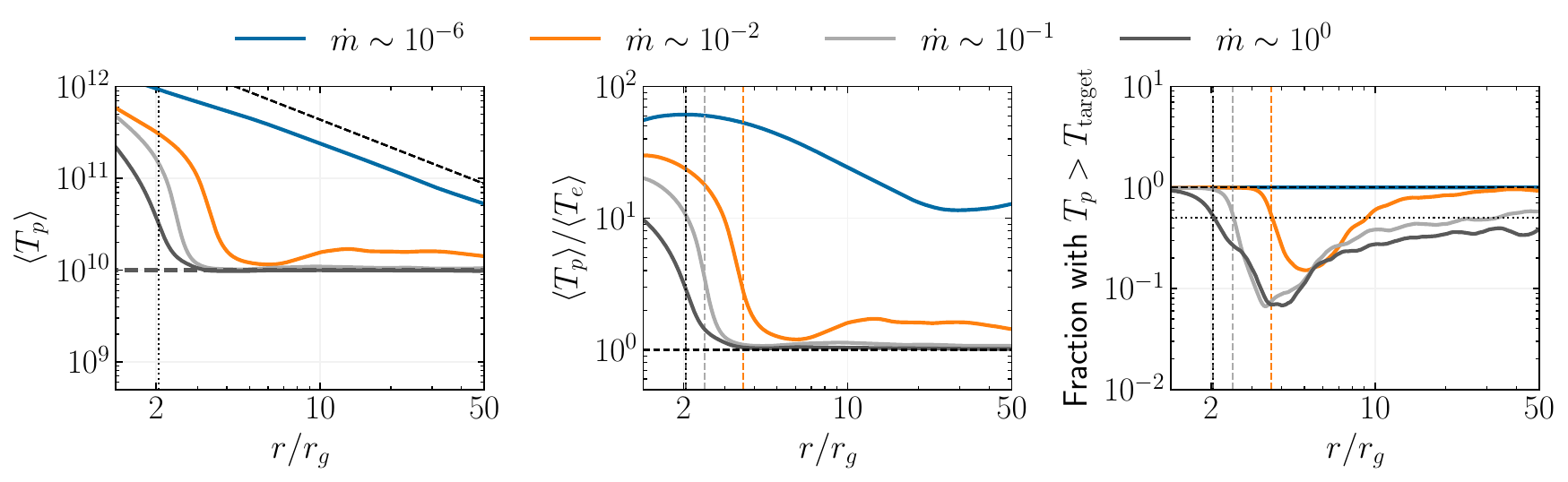}
    \caption{Thermodynamic properties of accretion flows as a function of accretion rate, showing the impact of Coulomb collisions. Left panel: the proton temperature, which determines the disc structure, as a function of radius. Black horizontal and slanted dashed lines show the target electron temperature and virial temperature, respectively. Center panel: the proton-to-electron temperature ratio, demonstrating the single-temperature vs. two-temperature regimes of the simulations. Dashed black horizontal line shows the single-temperature regime where $T_p=T_e$. Right panel: profile of the gas fraction above the target temperature. Black dotted horizontal line shows the maximum fraction of 1. Colored vertical dashed lines show the location of the transition radius (Eq.~\ref{eq:transition-radius}). Data are time-averaged over the last $2000r_g/c$.}
    \label{fig:thermo}
\end{figure*}

Coulomb collisions at high accretion rates strongly affect the radial structure of an accretion disc and change the thermodynamics of the disc (Fig.~\ref{fig:thermo}).
At low accretion rates, the canonical thick, two-temperature flow persists despite electrons having a temperature of $10^{10}~{\rm K}$ because Coulomb collisions are not efficient enough to cool down the protons (Fig.~\ref{fig:thermo} left).
Increasing the accretion rate to \highestM~enables stronger Coulomb collisions, which drive the M0 accretion disc to the single-temperature regime for $r\gtrsim r_{\rm ISCO}$.
In the intermediate regime of $\dot m\sim10^{-2}$, Coulomb collisions are only efficient in the outer parts of the disc and the proton temperature increases above the target electron temperature for $r\lesssim 4r_g$.

The temperature ratio $T_p/T_e$'s radial profile also depends on accretion rate.
For the hot accretion flow M6, the disc is two-temperature for all radii, with $T_p/T_e>100$.
In contrast, the highest accretion rate simulation M0 remains single-temperature until $r\approx r_{\rm ISCO}$, where decoupling between protons and electrons means $T_p/T_e$ reaches 100 at the event horizon.
For $\dot m\lesssim 1$, the ratio increases, reaching 10 at the ISCO and 30 at the event horizon for M2. 
For this simulation, the temperature ratio outside the ISCO is closer to $2$, suggesting that the upper layers of the disc are two-temperature while the midplane is single-temperature (see Sec.~\ref{sssec:vertical}).

To quantify these phases, we define a ``transition radius'' where more than 50\% of the gas at that radius has a gas temperature above the target temperature, similar to~\citet{hogg2018}:
\begin{equation}
    \left(\frac{T_p>T_{e, {\rm target}}}{T_p}\right)(r_{\rm tr})=0.5.\label{eq:transition-radius}
\end{equation}
The radial profile of gas with temperature above the target is shown in Fig.~\ref{fig:thermo}'s right panel. 
The M0 simulation has a transition radius $r_{\rm tr}=2.04r_g$, only slightly larger than the ISCO.
On the other extreme, the M6 simulation does not transition within the inflow equilibrium. 
The transition radius increases monotonically with accretion rate, with a value of $2.5r_g$ for M1 and $3.6r_g$ for M2.
The transition radius is plotted as a vertical dashed line with the color corresponding to the accretion rate.

\begin{figure*}
    \centering
    \includegraphics[width=\linewidth]{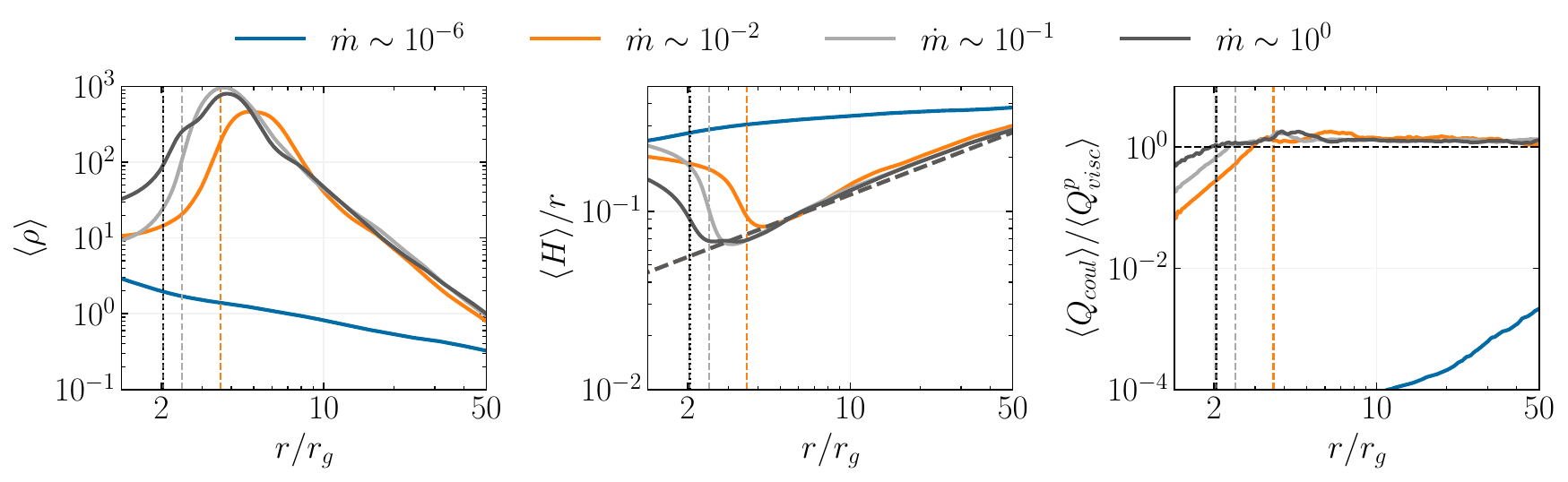}
    \caption{Structural properties of accretion flows as a function of accretion rate, showing the impact of Coulomb collisions. Vertical black dotted lines show the location of the ISCO. Left panel: the mass density radial profile, with the black dashed line showing the thin disc $\rho\sim r^{-3/2}$ scaling. Center panel: the disc scale height ratio $H/r$ (Eq.~\ref{eq:h-over-r}), with black dashed line showing the analytic value of $H/r$ for a constant gas temperature. Right panel: the ratio of Coulomb to viscous heating rates, showing the relative (sub)dominance of Coulomb collisions for different accretion rates. The horizontal black dashed line shows where the energy exchange rates are equal.  Colored vertical dashed lines show the location of the transition radius (Eq.~\ref{eq:transition-radius}). Time-averaged over the last $2000r_g/c$.}
    \label{fig:structure}
\end{figure*}

\subsubsection{Structure} \label{sssec:structure}
The thermodynamics of the accretion discs affect their global structure, as illustrated in Fig.~\ref{fig:structure}.
The low accretion rate simulation M6 displays a shallower density profile than the higher accretion rate simulations M0, M1, and M2. 
As expected, the density drops rapidly at the disc edge just outside the ISCO, though M2's density begins dropping at larger radii, $r\sim4r_g$ rather than $r\sim3r_g$, closer to the transition radius.

The change in density affects the disc scale height ratio $H/r$, as seen in Fig.~\ref{fig:structure}'s center panel.
Whereas the low accretion rate simulation M6 has a large, relatively constant ratio of $H/r\sim 0.3$, M0 and M1 follow the canonical thin disc scaling for a constant gas temperature at $r\gtrsim r_{\rm ISCO}$.
The intermediate accretion rate simulation M2 follows the thin disc scaling $H/r\sim r^{1/2}$ for constant gas temperature for $r\gtrsim4r_g$, but deviates significantly even outside the ISCO.
All the thin disc simulations show a significant increase in the disc scale height within the ISCO: M1 and M2 reach $H/r\sim0.2$ at the ISCO, a factor of about 4 greater than the targeted value.
For these simulations, the deviations from the thin disc $H/r$ occur for $r\lesssim r_{\rm tr}$.

To understand the origin of the change in radial disc structure, Fig.~\ref{fig:structure}'s right panel plots the density-weighted shell averaged values of the Coulomb energy exchange rate $Q_{\rm coul}$ to the dissipated viscous energy rate $Q_{\rm visc}^p$ received by the protons. 
Because Coulomb collisions cool the protons whereas viscous dissipation heats the protons, the ratio of the two quantities shows where we might expect the transition radius to occur. 
For M6, $Q_{\rm Coul}/Q_{\rm visc}^p\ll1$ as expected, since Coulomb collisions are subdominant in a hot accretion flow. 
For the thin disc simulations, the ratio reaches an equilibrium $Q_{\rm Coul}=Q_{\rm visc}^p$ for most radii.
Deviations again occur for $r<r_{\rm tr}$.

\subsubsection{Timescales} \label{sssec:timescales}
\begin{figure*}
    \centering
    \includegraphics[width=\linewidth]{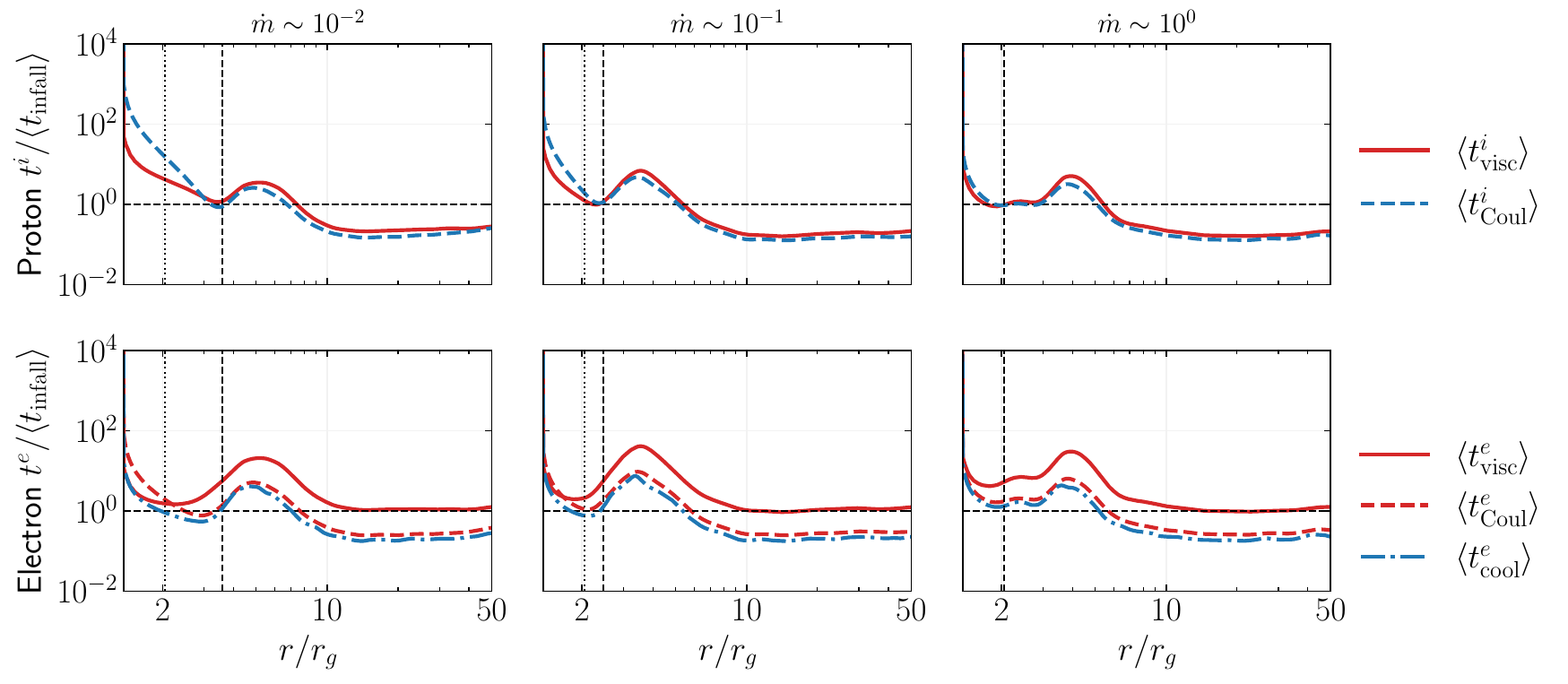}
    \caption{Demonstration of the proton and electron timescales in the system's relation to the infall time depending on accretion rate and radius. Columns show the accretion rates from lowest (left) to highest (right), simulations M2, M1, M0 in order. Rows show proton timescales (top) or electron timescales (bottom). In all panels, the dashed horizontal black line shows the infall time. Red indicates that the physical process heats the indicated species, while processes in blue cool the indicated species. Vertical dotted line shows the location of the ISCO, while vertical dashed lines show the transition radius (Eq.~\ref{eq:transition-radius}). Time-averaged over the last $2000r_g/c$.}
    \label{fig:timescales}
\end{figure*}

To understand the origin of the thermodynamics and structure presented in Figs.~\ref{fig:thermo} and~\ref{fig:structure}, Fig.~\ref{fig:timescales} plots the relevant timescales in the system (Eqns.~\ref{eq:tCoul} and~\ref{eq:tVisc}).
These timescales are normalized to the infall time (Eq.~\ref{eq:tinfall}) such that lines above 1 show physical processes that cannot operate before the gas accretes.
The timescales are separated into proton (top) and electron (bottom) to demonstrate how the electron cooling function affects the gas structure by influencing protons through Coulomb collisions. 

For the high accretion rate simulation M0 (Fig.~\ref{fig:timescales} right column), the thermodynamic timescales are shorter than the infall time for all radii except $r<4r_g$.
These short timescales indicate that the viscously-dissipated energy is immediately ``radiated'', i.e. removed through the electron-only cooling function, in agreement with the assumptions made in~\citet{ss73}. 
As the bottom panel shows, the electron heating for $r>r_{\rm ISCO}$ is dominated by Coulomb collisions rather than viscous heating.
Within the ISCO, all relevant timescales become greater than the infall time. 
Notably, the proton heating timescale remains smaller than the Coulomb cooling timescale, showing that protons increase in temperature within the ISCO due to viscous heating, leading to the thermodynamic and structural properties noted earlier that show $H/r$ within the ISCO increasing above expected thin disc values.

At more intermediate accretion rates M1 and M2, the increase in heating/cooling timescales above the infall time occurs at the transition radius rather than the ISCO (Fig.~\ref{fig:timescales} center/left columns). 
The decoupling at larger radii means that the properties of the disc diverge more by the time the accreting gas reaches the event horizon. 
Indeed, for M2 the proton Coulomb timescale becomes two orders of magnitude longer than the proton heating timescale within the transition radius, leading the a more ADAF-like disc structure.
Similar to the M1 and M0 simulations, the electron heating at radii larger than the transition radius is dominated by Coulomb collisions rather than viscous heating.
Within the ISCO, however, the Coulomb heating time for electrons increases faster than any other timescale, likely due to its dependence on the mass density, which drops by two orders of magnitude.
Electron cooling still successfully balances the viscous heating within the ISCO, as seen by the overlap of $t_{\rm cool}^e$ and $t_{\rm heat}^e$.

% ---------------------------------------------
\subsection{Radiative Properties as a Function of Accretion Rate} \label{ssec:results-over-mdot}
Including Coulomb collisions can affect the amount and distribution of radiation coming from the accretion disc.
In this section, we will compare to the canonical Novikov-Thorne model and explore how much cooling occurs in two-temperature, optically-thin regions of the accretion flow.

\subsubsection{Radial Distribution of Luminosity and Radiative Efficiency}
\begin{figure}
    \centering
    \includegraphics[width=0.9\linewidth]{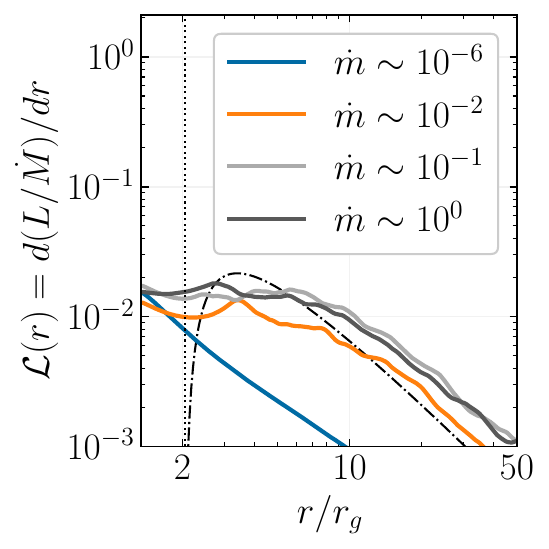}
    \caption{Radial profile of the luminosity per unit accretion rate's derivative $\mathcal{L}(r)$ (Eq.~\ref{eq:dLdr}). The black dash-dot line shows the analytic ~\citet{novikovthorne} prediction for $a=0.9375$.
    Vertical dotted line shows the location of the ISCO. 
    Time-averaged over the last $2000r_g/c$.}
    \label{fig:dLdr}
\end{figure}
This section tests the basic principles of viscous-based accretion and explores how Coulomb decoupling affects the resulting radiative properties.
The Novikov-Thorne model assumes that 100\% of the dissipated energy converts into radiation locally, resulting in large radiative efficiencies~\citep{novikovthorne,page1974}.
To probe where the luminosity comes from, we calculate the radial derivative of the luminosity, $\mathcal{L}(r)$ (Eq.~\ref{eq:dLdr}), shown in Fig.~\ref{fig:dLdr}.
The intermediate to high accretion rate simulations M2, M1, and M0 deviate from the Novikov-Thorne result (black dash-dot line) for radii $r\lesssim10r_g$.
However, whereas the Novikov-Thorne result famously decreases to zero at the ISCO, these simulations all emit a significant fraction of their luminosity within the ISCO.
Indeed, $\mathcal{L}$ mostly flattens off within the ISCO instead of disappearing or even decreasing.

\begin{figure}
    \centering
    \includegraphics[width=0.9\linewidth]{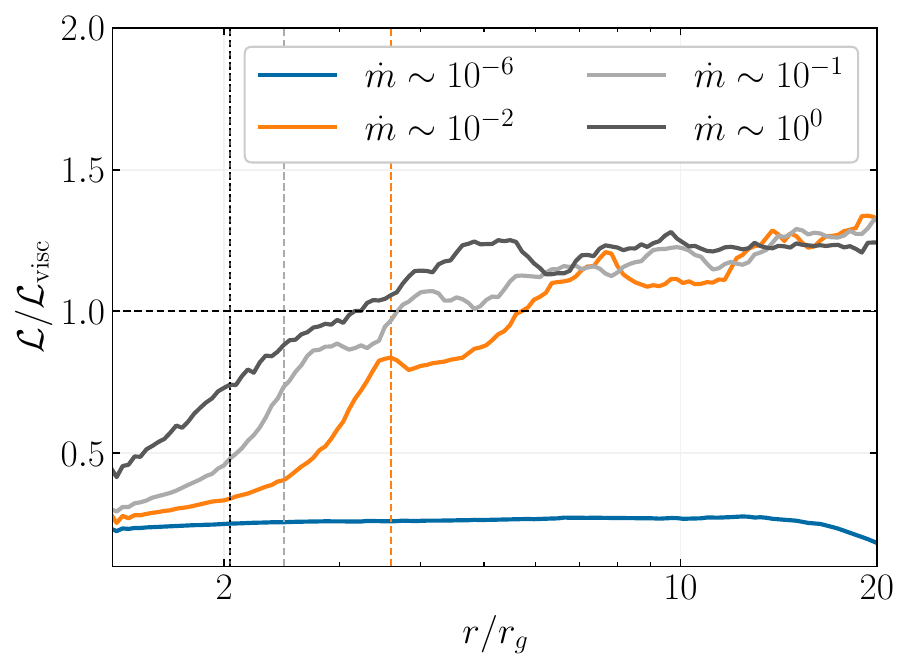}
    \caption{Comparison of $\mathcal{L}/\mathcal{L}_{\rm visc}$ at each accretion rate over radius. The horizontal dashed black line shows the isothermal single-temperature limit of $\mathcal{L}/\mathcal{L}_{\rm visc}\to1$. Vertical black dotted line shows the location of the ISCO; vertical dashed lines show the transition radius for each simulation (Eq.~\ref{eq:transition-radius}). Time-averaged over the last $2000r_g/c$.}
    \label{fig:QcoolOverQvisc}
\end{figure}

The accretion flows become increasingly radiatively inefficient at smaller radii, as seen in Fig.~\ref{fig:QcoolOverQvisc}.
To compare with the isothermal single-temperature regime, we plot the ratio of the shell-integrated luminosity from electron cooling to the luminosity from total viscous heating $\mathcal{L}/\mathcal{L}_{\rm visc}$. 
In the instantaneous cooling (isothermal) single-temperature regime, this ratio is always 1, though with a finite cooling time this ratio can deviate from 1.
As Fig.~\ref{fig:QcoolOverQvisc} shows, the M6 simulation has a fraction $\mathcal{L}/\mathcal{L}_{\rm visc}\sim0.3<1$ for all radii, which is set by the electron heating prescription for $\delta_e$.
By comparison, the ratio for the higher accretion rate simulations increases monotonically with radius and with accretion rate.
The local radiative efficiency for M0 reaches 0.5 at its lowest value, whereas M1 and M2 approach the M6 value at the event horizon. 
The three higher accretion rate simulations approach $\mathcal{L}/\mathcal{L}_{\rm visc}\to 1$ at about twice their respective $r_{\rm tr}$.

\begin{table}
\centering
\begin{tabular}{|l | c|c| c}
 & $L_\infty^{\rm visc}/\dot M c^2$ & $L_\infty/\dot M c^2$ & $L_\infty/L_\infty^{\rm visc}$\\ \hline\hline
M6 (\lowestM) & 0.17  & 0.04 & 0.21 \\ \hline
M2 ($\dot m\sim 10^{-2})$ & 0.17 & 0.16 & 0.95 \\ \hline
M1 ($\dot m\sim 10^{-1})$ & 0.26  & 0.29 & 1.12 \\ \hline
M0 (\highestM) & 0.25 &  0.30& 1.20\\ \hline
\end{tabular} 
\caption{Efficiency fractions (see Eq.~\ref{eq:integrated-L}) time-averaged over the last $2000r_g/c$ and over $r_{\rm EH}<r<20r_g$.}\label{tab:efficiency}
\end{table}

The volume-integrated viscous and cooling energy rates are displayed in Table~\ref{tab:efficiency}.
The viscous dissipation rate $L_\infty^{\rm visc}/L_\infty$ is best compared to the Novikov-Thorne value of $0.18$ for a spin of 0.9375. 
The lower accretion rate simulations M6 and M2 in particular have a viscous dissipation rate $\eta_{\rm visc}\equiv Q_{\rm visc}/\dot M c^2$ almost precisely equal to the Novikov-Thorne value.
The higher accretion rate simulations M1 and M0 have viscous efficiencies larger than the Novikov-Thorne value, presumably due to dissipation within the plunging region.

The ratio $L_\infty/\dot Mc^2$, which can be considered the equivalent of radiative efficiency for simulations with a cooling function, depends strongly on accretion rate.
As expected, the radiatively-inefficient simulation M6 converts only a small fraction ($4\%$) of the viscous heating into radiation.
This fraction is notably higher than previous two-temperature, weakly magnetized simulations, which found a radiative efficiency of $2\times10^{-3}$ for $\dot m\sim10^{-6}$~\citep{ryan2017}.
The difference likely occurs due to the assumed fast electron cooling in the electron cooling function.
M6 therefore provides an upper limit for the radiative efficiency at low accretion rates.
The M2 simulation has a radiative efficiency $\eta=0.17$, consistent with the Novikov-Thorne value.
The higher accretion rate simulations 
M1 and M0's radiative efficiencies exceed $\eta_{NT}$, with $\eta=0.29$ and $0.30$, respectively.
Slightly more than 100\% of the viscous energy is dissipated into radiation for M1 and M0, with $L_\infty/L_\infty^{\rm visc}=1.12$ and 1.2, respectively, implying another source of heating.
If adiabatic heating accounts for the heating that leads to more cooling, it suggests that the emission comes from the inner regions of the flow where adiabatic compression is strongest, i.e. within the plunging region, as previously shown in Fig.~\ref{fig:dLdr}.

\subsubsection{Radiation from Two-temperature, Optically-Thin Regions} \label{sssec:twotemp}
\begin{figure*}
    \centering
    \includegraphics[width=0.95\linewidth]{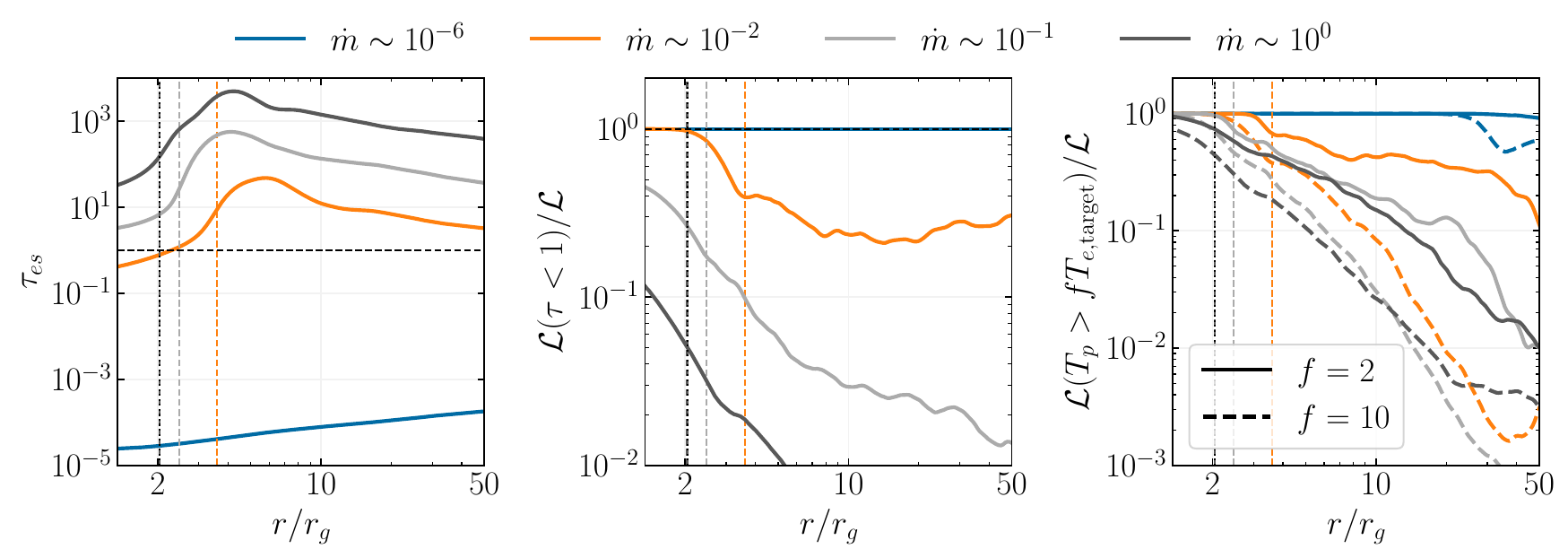}
    \caption{Radiative properties of thin accretion flows as a function of accretion rate. 
    Left panel: the optical depth at the midplane (Eq.~\ref{eq:opt-depth}), with black dashed line showing where the disc becomes optically thick at $\tau=1$. 
    Center panel: fraction of shell-integrated luminosity emitted from optically-thin regions (see Eq.~\ref{eq:dLdr}), with horizontal black dashed line showing where 100\% of the emission comes from optically-thin regions.
    Right panel: the fraction of shell-integrated luminosity from regions with $T_p/T_e>f$, where $f=2$ is shown by solid lines and $f=10$ is shown with dashed lines.
    In all panels, the vertical dotted line shows the location of the ISCO, while vertical dashed lines show the transition radius (Eq.~\ref{eq:transition-radius}). Time-averaged over the last $2000r_g/c$.}
    \label{fig:dissipation}
\end{figure*}
The electron-only cooling function in these two-temperature simulations allows probing how much of the energy lost to cooling comes from two-temperature and optically-thin regions.
Fig.~\ref{fig:dissipation}'s left panel shows the midplane optical depth (Eq.~\ref{eq:opt-depth}) for different accretion rates. 
While the lowest accretion rate simulation M6 is optically-thin at all radii as expected, the canonical thin disc M2 drops two orders of magnitude in optical depth between just outside the ISCO and just outside the event horizon.
The higher accretion rate simulations are marginally optically thick within the ISCO, with $\tau\sim10$ for M1. 

A significant fraction of electron cooling takes place in optically-thin regions.
As Fig.~\ref{fig:dissipation}'s center panel shows, at intermediate accretion rates (M2), approximately 20\% of light at large radii comes from optically-thin regions.
This fraction increases to 100\% within the ISCO, where the flow becomes optically thin everywhere. 
For higher accretion rates, the fraction of cooling in optically-thin regions drops. 
Within the ISCO, the highest accretion rate simulation M0 has close to 10\% of radiation coming from optically-thin regions. 
We note that if this ratio instead considers cooling from within marginally optically-thick regions, i.e. with $\tau<10$, the amount of cooling increases greatly, to $>30\%$ in the M0 simulation (not shown).

Fig.~\ref{fig:dissipation}'s right panel plots the shell-integrated ratio of electron cooling that occurs in regions with proton temperatures more than twice the target electron temperature (solid lines) and more than ten times the target electron temperature (dashed lines).
Within the ISCO, close to all of the cooling originates from strongly two-temperature regions.
Only the M0 simulation ratio does not reach 1 at the ISCO, instead increasing from around 0.8 towards 1 at the event horizon. 
Even outside the ISCO, more than 10\% of the radiation within $10r_g$ comes from regions with $T_p>2T_{e, {\rm target}}$ for all accretion rates.

% --------------------------------------------
% --------------------------------------------
\section{Discussion} \label{sec:discussion}

\subsection{Disc Truncation Radius Depends on Accretion Rate}
\begin{figure}
    \centering
    \includegraphics[width=0.9\linewidth]{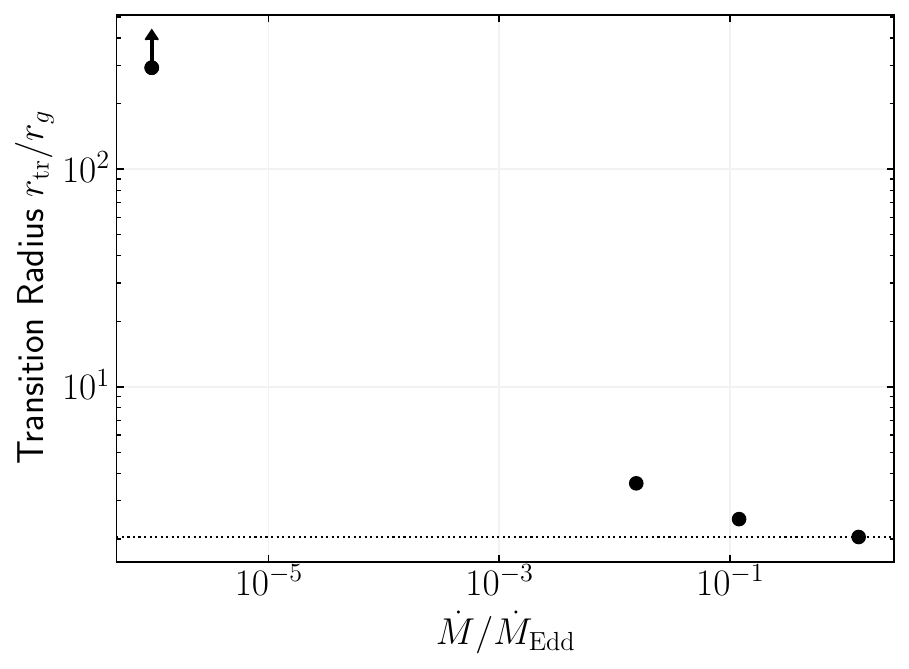}
    \caption{The transition radius (Eq.~\ref{eq:transition-radius}) as a function of accretion rate normalized to the Eddington accretion rate with nominal efficiency $\eta=0.1$. Black dotted line shows the location of the ISCO for the black hole spin $a=0.9375$.}
    \label{fig:rtr}
\end{figure}
With a suite of 3D GRMHD simulations of weakly magnetized thin accretion discs, we have demonstrated that Coulomb collisions alone can lead to the formation of a ``truncated'' disc, wherein the single-temperature accretion disc transitions to a hot, two-temperature flow within some radius.
These simulations show that the transition radius decreases monotonically with accretion rate from $\sim5r_g$ at intermediate accretion rates to $r_{\rm ISCO}$ at high accretion rates (Fig.~\ref{fig:rtr}).
The smallest accretion rate simulation does not truncate within the inflow equilibrium radius, and therefore the ``transition'' is marked by a lower limit.
Outside the transition radius, the flow reaches the canonical single-temperature regime. 

Although the target temperature in these simulations is larger than expected from Compton scattering, the presence of the decoupled flow over a factor of 100 in accretion rate suggests that the truncation is a robust result. 
The exact accretion rates listed in these simulations may not remain the same with a lower target temperature since the proton temperature and thus density would also be affected by a change in target temperature.
However, the presence of the truncation over such a large range in accretion rate suggests that models with more accurate temperatures would also result in truncation, at least within the ISCO.
A simulation with a target temperature lowered by a factor of 100, yielding a density a factor of approximately 10 higher, will still decouple within the ISCO, as shown analytically~\citep{hankla2022}.
Future work will examine more realistic target temperatures, including prescribing $H/r$ as constant (with $T_{e, \rm target}\propto T^{1/2}$) as is the case in radiation-pressure-dominated discs.

\subsubsection{Implications for XRB State Transitions}
The spectral state transitions of X-ray Binaries have long been proposed to occur through the disc truncation model~\citep{esin1997, done2007}.
Coulomb collisions often cause the truncation in these models, with the transition radius predicted to occur at hundreds of gravitational radii rather than within 10$r_g$ as seen in M2 and at higher accretion rates. 
Although it is possible that accretion rates between M2 and M6 will truncate at larger radii, probing a transition at those large scales will require evolving the simulations for many dynamical times at those radii. 

Notably, the disc truncation presented in this simulation suite occurs due to Coulomb collisions where gas pressure dominates the disc structure, namely the disc does not become magnetically arrested (MAD). 
In accretion discs with stronger magnetic fields, the magnetic fields likely play a significant role in truncating the disc~\citep{liska2022}, although the radius that the disc observationally truncates could occur at some location within the MAD radius~\citep{scepi2024b}.
XRB state transitions are likely governed by a combination of accretion rate changes and magnetic field advection~\citep{begelman2014}. 

\subsubsection{Implications For Particle Acceleration within the ISCO}
The truncation of a thin disc into a two-temperature flow at high accretion rates has implications for nonthermal particle acceleration within the ISCO.
Because of the drop in density within the ISCO by two orders of magnitude, the collision times between electrons also drop, as seen in the Coulomb exchange time $t^{e,i}_{\rm Coul}$ in Figs.~\ref{fig:timescales-demo} and~\ref{fig:timescales} and calculated analytically in~\citet{hankla2022}.
These conditions could be sufficient to produce the hard X-ray tail $>10~{\rm keV}$ observed in the soft state of XRBs~\citep{mcconnell2002}.

\subsection{Thin Discs Have Two-temperature Regions}
The two-temperature nature of this inner accretion flow modifies both the amount of observed radiation and the spectral shape of the emitted radiation.
Since a significant fraction of the emission comes from regions with $T_p>2T_e$ (Fig.~\ref{fig:dissipation} right), these regions could significantly impact the total spectra observed from these systems.
In particular, this decoupling also suggests that the emission from the flow cannot be modeled as a blackbody in thermal equilibrium.
Instead, the light from these regions is likely optically-thin and nonthermal.
The optically-thin nature could have implications for produces the hard spectrum of XRBs through multiple Compton scatterings, especially if the radial optical depth is large~\citep{dexter2021}. 
This lack of thermal equilibrium could impact models that rely on thermal emission for measuring black hole spin; see~\citet{reynolds2021}.

\subsubsection{Comparison to Single-Temperature Thin Discs}
The increase in luminosity within the ISCO compared to the Novikov-Thorne model was predicted with the inclusion of large-scale magnetic torques at the ISCO~\citep{gammie1999,agol2000}.
Single-temperature simulations of thin discs have previously seen an excess luminosity above the Novikov-Thorne zero-emission limit~\citep{penna2010} inside $\lesssim2r_{\rm ISCO}$.
Single-temperature strongly-magnetized discs have seen flux profiles about an order of magnitude above the Novikov-Thorne limit at its peak~\citep{dhang2025}. 
Although the total radiative efficiency from single-temperature simulations with a cooling function exceeds the Novikov-Thorne limit by $\sim80\%$~\citep{avara2016}, this number decreases with M1 radiation treatment~\citep{morales2018}.
As Fig.~\ref{fig:structure}'s right panel shows, including two-temperature physics allows the electrons to decouple from protons, resulting in a significant reduction in the light received from the inner hot flow by a factor of 2 or more (Fig.~\ref{fig:QcoolOverQvisc}).
This lower radiative efficiency could affect the importance of emission within the plunging region on black hole spin modelling.

\begin{figure*}
    \centering
    \includegraphics[width=\linewidth]{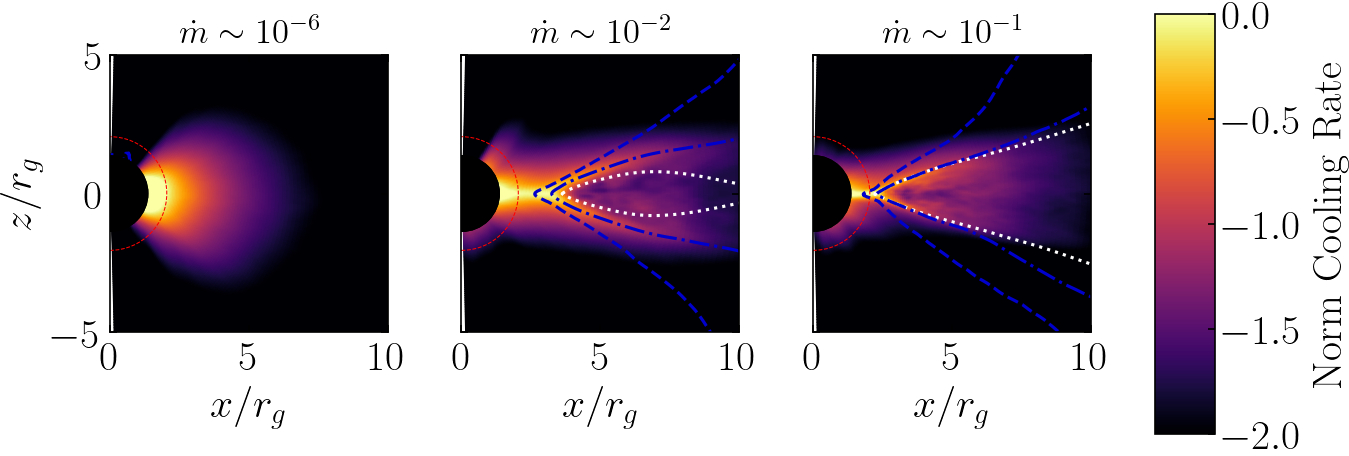}
    \caption{Vertical slices of the azimuthally-averaged cooling rate $-u_tQ_{\rm cool}$ for different accretion rates, normalized to the midplane value at the ISCO. The white dotted contour shows $\tau_{\rm es}=10$, while the dark blue dash-dot and dashed lines show $T_p/T_e=2$ and 10, respectively. The red dashed line shows the equatorial ISCO radius. Time-averaged over the last $2000r_g/c$.}
    \label{fig:slices}
\end{figure*}
\subsection{The Origin of the Corona}
The hard spectral state of XRBs suggests the presence of hot electrons that Compton upscatter photons to hard X-rays.
The properties of and spatial distribution of these hot electrons, collectively termed ``the corona'', can affect black hole spin measurements and interpretations of X-ray polarimetry with telescopes such as IXPE~\citep{ixpe}. 
Although spectral-timing and microlensing techniques point towards a compact corona within $\lesssim10r_g$~\citep{zoghbi2011, uttley2014, chartas2009}, the exact origin and structure of the corona has remained a mystery; see~\citet{laha2025} for an observational review of the corona in radio-quiet AGN. 
In this section, we explore the nature of the two-temperature ``coronal regions'' in these thin disc simulations.

\subsubsection{Vertical Extent of the Two-Temperature Regions} \label{sssec:vertical}

Fig.~\ref{fig:slices} shows the distribution of luminosity from the electron cooling function.
Notably, a significant amount of cooling originates from above the disc out to $\sim10r_g$ in a sandwich-esque configuration. 
Within the truncated region, the emission clusters close to the equatorial plane rather than becoming perfectly spherical.
We again note that the emission outside the contour for $T_p/T_e=10$ is likely highly nonthermal. 
The amount of radiation from above the disc body is similar to the amount found in single-temperature simulations of thin, strongly magnetized accretion discs~\citep{scepi2024} and local, two-temperature simulations~\citep{bambic2024}.
These findings have implications for the shape of the X-ray corona as probed by spectral-timing studies and X-ray polarimetry.

\subsection{Impact of Two-temperature Physics on Black Hole Spin Measurements}
A common method to measure black hole spin involves identifying the disc peak temperature and mapping that temperature to a specific radius related to the ISCO, which then provides the black hole spin; see~\citet{reynolds2021}.
This thermal continuum fitting model is therefore extremely sensitive to the location of the maximum temperature.
Two-temperature physics impacts the location of maximum temperature in two main ways.

First, at intermediate accretion rates, Coulomb decoupling leads to the transition away from a thin disc at radii larger than the ISCO (e.g. Fig.~\ref{fig:rtr}).
For example, the M2 simulation's ISCO would be observationally identified as having an ISCO around $4r_g$, rather than the true value of $\approx2r_g$ for $a=0.9375$.
Thus truncation due to two-temperature effects could lead to underestimating black hole spin for intermediate accretion rates.

Second, two-temperature effects reduce the impact of thermal emission from within the ISCO compared to single-temperature models.
Single-temperature models have long demonstrated that thermal emission from within the ISCO can lead to misidentification of the ISCO location, thereby overestimating black hole spin~\citep{penna2010}.
As we show in Sec.~\ref{ssec:results-over-mdot}, including two-temperature effects reduces the efficiency of thermal emission from within the ISCO.
Because Coulomb collisions become inefficient, protons heat the electrons less, leading to less electron cooling than in single-temperature models~\citep[e.g.]{mummery2024}.
At the highest accretion rate, the thermal emission at the ISCO is $\approx75\%$ of the single-temperature prediction, dropping to $<50\%$ for lower accretion rates
(Fig.~\ref{fig:QcoolOverQvisc}).
In addition, two-temperature physics shows that a significant amount of radiation from within the ISCO is optically thin (Fig.~\ref{fig:dissipation} central panel) rather than thermal.
Emission from the plunging region therefore likely contains a significant nonthermal component rather than appearing purely thermal, as predicted analytically~\citep{hankla2022} and suggested by detailed postprocessing of GRMHD simulations~\citep{zhu2012}.

% ------------------------------------------------
\subsection{Numerical Considerations} \label{ssec:other}
The electron-only cooling function presented in this paper is by necessity a simplification of the ``true'' physics of a two-temperature accretion disc. 
This model does not include non-local effects due to Compton scattering, which could affect the truncation structure~\citep{esin1997b}.
In addition, it assumes that the main coupling mechanism between protons and electrons is due to Coulomb collisions.
While this assumption likely holds for the conditions in an accretion disc, it is in principle possible for other coupling mechanisms to exist~\citep{begelman1988}.
The form of the cooling function assumes a cooling time proportional to the orbital time, whereas the physical cooling time is likely much shorter, resulting in an electron temperature that deviates from the target temperature within the ISCO. 

The cooling function also by nature assumes optically-thin radiation, since the radiation energy is simply removed from the gas without interacting with any gas along the line-of-sight.
Presumably this assumption will break down at higher accretion rates where the disc becomes dominated by radiation pressure and in optically-thick regions.
Because radiation pressure changes the disc temperature and density profiles, future work will address disc truncation in the radiation pressure dominated regime with a constant $H/r$ rather than a constant target temperature, as was motivated by considering Compton scattering as the dominant cooling mechanism.
More accurate treatment of radiation remains computationally expensive, and requires circumventing the thermal instability, whose impact on global disks remains unclear~\citep{mishra2016, fragile2018, mishra2022}.
The numerically tractable choice of a cooling function provides a minimal model for capturing the effects of radiation on a stable disc and permits studying the impact of radiation as a function of accretion rate. 

In addition, computational constraints limited the numerical resolution of these simulations.
For the weakly magnetized discs presented here, the low numerical resolution as demonstrated by the quality factors in Table~\ref{tab:last-times} potentially suppressed some turbulence due to the MRI. 
However, since the temporal evolution of the accretion rate and viscous dissipation rate was roughly constant over the times considered (not shown), this effect likely does not change the conclusions drawn in this paper.

\section{Conclusions}
By treating radiation through an electron cooling function and allowing protons to cool exclusively through Coulomb coupling with electrons, we have probed the radial and vertical structure of thin accretion discs' truncation to hot accretion flows.
We find that the plunging region of thin accretion discs is necessarily two-temperature.
This two-temperature region can extend to $r>r_{\rm ISCO}$ for intermediate accretion rates, forming an inner ``truncated'' disc whose radial location depends on accretion rate, with implications for XRB state transition and spin measurements.
The emission from this two-temperature region is likely optically-thin and possibly nonthermal, and therefore its emission cannot be adequately captured with MHD simulations.

Future work will continue to explore this radial transition from thin accretion disc to two-temperature flow. 
In particular, the dependence of the transition radius on the electron target temperature should be explored.
Reducing the target temperature  to $10^9$ K will require higher numerical resolution. 
Changing the radial dependence of the target temperature to, for example, yield a constant $H/r$, as in the case for radiation-pressure-dominated discs, will also be interesting. 
Black hole spin could also affect the transition radius, since the location of the ISCO depends monotonically on spin.

\section*{Acknowledgements}
Support for this work was provided by NASA through the NASA Hubble Fellowship grant \#HF2-51555 awarded by the Space Telescope Science Institute, which is operated by the Association of Universities for Research in Astronomy, Inc., for NASA, under contract NAS5-26555.
This work was supported in part by NASA Astrophysics Theory Program grants 80NSSC20K0527 and 80NSSC24K1094, by a \texttt{Chandra} grant TM3-24003X, by an Alfred P. Sloan Research Fellowship (JD), and by the National Science Foundation Graduate Research Fellowship Program under Grant No. DGE 1650115 (AH).
The authors thank B. Ryan, J. Miller, and A. Dittmann for helpful discussions.

%%%%%%%%%%%%%%%%%%%%%%%%%%%%%%%%%%%%%%%%%%%%%%%%%%
\section*{Data Availability}
The simulation data underlying this article will be shared on reasonable request to the corresponding author.

%%%%%%%%%%%%%%%%%%%% REFERENCES %%%%%%%%%%%%%%%%%%

% The best way to enter references is to use BibTeX:
\bibliographystyle{mnras}
\bibliography{refs} % if your bibtex file is called example.bib

\begin{thebibliography}{}
\makeatletter
\relax
\def\mn@urlcharsother{\let\do\@makeother \do\$\do\&\do\#\do\^\do\_\do\%\do\~}
\def\mn@doi{\begingroup\mn@urlcharsother \@ifnextchar [ {\mn@doi@} {\mn@doi@[]}}
\def\mn@doi@[#1]#2{\def\@tempa{#1}\ifx\@tempa\@empty \href {http://dx.doi.org/#2} {doi:#2}\else \href {http://dx.doi.org/#2} {#1}\fi \endgroup}
\def\mn@eprint#1#2{\mn@eprint@#1:#2::\@nil}
\def\mn@eprint@arXiv#1{\href {http://arxiv.org/abs/#1} {{\tt arXiv:#1}}}
\def\mn@eprint@dblp#1{\href {http://dblp.uni-trier.de/rec/bibtex/#1.xml} {dblp:#1}}
\def\mn@eprint@#1:#2:#3:#4\@nil{\def\@tempa {#1}\def\@tempb {#2}\def\@tempc {#3}\ifx \@tempc \@empty \let \@tempc \@tempb \let \@tempb \@tempa \fi \ifx \@tempb \@empty \def\@tempb {arXiv}\fi \@ifundefined {mn@eprint@\@tempb}{\@tempb:\@tempc}{\expandafter \expandafter \csname mn@eprint@\@tempb\endcsname \expandafter{\@tempc}}}

\bibitem[\protect\citeauthoryear{{Agol} \& {Krolik}}{{Agol} \& {Krolik}}{2000}]{agol2000}
{Agol} E.,  {Krolik} J.~H.,  2000, \mn@doi [\apj] {10.1086/308177}, \href {https://ui.adsabs.harvard.edu/abs/2000ApJ...528..161A} {528, 161}

\bibitem[\protect\citeauthoryear{{Avara}, {McKinney}  \& {Reynolds}}{{Avara} et~al.}{2016}]{avara2016}
{Avara} M.~J.,  {McKinney} J.~C.,   {Reynolds} C.~S.,  2016, \mn@doi [\mnras] {10.1093/mnras/stw1643}, \href {https://ui.adsabs.harvard.edu/abs/2016MNRAS.462..636A} {462, 636}

\bibitem[\protect\citeauthoryear{{Bambic}, {Quataert}  \& {Kunz}}{{Bambic} et~al.}{2024}]{bambic2024}
{Bambic} C.~J.,  {Quataert} E.,   {Kunz} M.~W.,  2024, \mn@doi [\mnras] {10.1093/mnras/stad3261}, \href {https://ui.adsabs.harvard.edu/abs/2024MNRAS.527.2895B} {527, 2895}

\bibitem[\protect\citeauthoryear{{Begelman} \& {Armitage}}{{Begelman} \& {Armitage}}{2014}]{begelman2014}
{Begelman} M.~C.,  {Armitage} P.~J.,  2014, \mn@doi [\apjl] {10.1088/2041-8205/782/2/L18}, \href {https://ui.adsabs.harvard.edu/abs/2014ApJ...782L..18B} {782, L18}

\bibitem[\protect\citeauthoryear{{Begelman} \& {Chiueh}}{{Begelman} \& {Chiueh}}{1988}]{begelman1988}
{Begelman} M.~C.,  {Chiueh} T.,  1988, \mn@doi [\apj] {10.1086/166698}, \href {https://ui.adsabs.harvard.edu/abs/1988ApJ...332..872B} {332, 872}

\bibitem[\protect\citeauthoryear{{Chartas}, {Kochanek}, {Dai}, {Poindexter}  \& {Garmire}}{{Chartas} et~al.}{2009}]{chartas2009}
{Chartas} G.,  {Kochanek} C.~S.,  {Dai} X.,  {Poindexter} S.,   {Garmire} G.,  2009, \mn@doi [\apj] {10.1088/0004-637X/693/1/174}, \href {https://ui.adsabs.harvard.edu/abs/2009ApJ...693..174C} {693, 174}

\bibitem[\protect\citeauthoryear{{Dexter}, {Scepi}  \& {Begelman}}{{Dexter} et~al.}{2021}]{dexter2021}
{Dexter} J.,  {Scepi} N.,   {Begelman} M.~C.,  2021, \mn@doi [\apjl] {10.3847/2041-8213/ac2608}, \href {https://ui.adsabs.harvard.edu/abs/2021ApJ...919L..20D} {919, L20}

\bibitem[\protect\citeauthoryear{{Dhang}, {Dexter}  \& {Begelman}}{{Dhang} et~al.}{2025}]{dhang2025}
{Dhang} P.,  {Dexter} J.,   {Begelman} M.~C.,  2025, \mn@doi [\apj] {10.3847/1538-4357/ada76e}, \href {https://ui.adsabs.harvard.edu/abs/2025ApJ...980..203D} {980, 203}

\bibitem[\protect\citeauthoryear{{Done}, {Gierli{\'n}ski}  \& {Kubota}}{{Done} et~al.}{2007}]{done2007}
{Done} C.,  {Gierli{\'n}ski} M.,   {Kubota} A.,  2007, \mn@doi [\aapr] {10.1007/s00159-007-0006-1}, \href {https://ui.adsabs.harvard.edu/abs/2007A&ARv..15....1D} {15, 1}

\bibitem[\protect\citeauthoryear{{Esin}}{{Esin}}{1997}]{esin1997b}
{Esin} A.~A.,  1997, \mn@doi [\apj] {10.1086/304129}, \href {https://ui.adsabs.harvard.edu/abs/1997ApJ...482..400E} {482, 400}

\bibitem[\protect\citeauthoryear{{Esin}, {McClintock}  \& {Narayan}}{{Esin} et~al.}{1997}]{esin1997}
{Esin} A.~A.,  {McClintock} J.~E.,   {Narayan} R.,  1997, \mn@doi [\apj] {10.1086/304829}, \href {https://ui.adsabs.harvard.edu/abs/1997ApJ...489..865E} {489, 865}

\bibitem[\protect\citeauthoryear{{Fabian} et~al.,}{{Fabian} et~al.}{2020}]{fabian2020}
{Fabian} A.~C.,  et~al., 2020, \mn@doi [\mnras] {10.1093/mnras/staa564}, \href {https://ui.adsabs.harvard.edu/abs/2020MNRAS.493.5389F} {493, 5389}

\bibitem[\protect\citeauthoryear{{Fishbone} \& {Moncrief}}{{Fishbone} \& {Moncrief}}{1976}]{fishbonemoncrief}
{Fishbone} L.~G.,  {Moncrief} V.,  1976, \mn@doi [\apj] {10.1086/154565}, \href {https://ui.adsabs.harvard.edu/abs/1976ApJ...207..962F} {207, 962}

\bibitem[\protect\citeauthoryear{{Fragile}, {Etheridge}, {Anninos}, {Mishra}  \& {Klu{\'z}niak}}{{Fragile} et~al.}{2018}]{fragile2018}
{Fragile} P.~C.,  {Etheridge} S.~M.,  {Anninos} P.,  {Mishra} B.,   {Klu{\'z}niak} W.,  2018, \mn@doi [\apj] {10.3847/1538-4357/aab788}, \href {https://ui.adsabs.harvard.edu/abs/2018ApJ...857....1F} {857, 1}

\bibitem[\protect\citeauthoryear{{Gammie}}{{Gammie}}{1999}]{gammie1999}
{Gammie} C.~F.,  1999, \mn@doi [\apjl] {10.1086/312207}, \href {https://ui.adsabs.harvard.edu/abs/1999ApJ...522L..57G} {522, L57}

\bibitem[\protect\citeauthoryear{{Gammie}, {McKinney}  \& {T{\'o}th}}{{Gammie} et~al.}{2003}]{gammie2003}
{Gammie} C.~F.,  {McKinney} J.~C.,   {T{\'o}th} G.,  2003, \mn@doi [\apj] {10.1086/374594}, \href {https://ui.adsabs.harvard.edu/abs/2003ApJ...589..444G} {589, 444}

\bibitem[\protect\citeauthoryear{{Hankla}, {Scepi}  \& {Dexter}}{{Hankla} et~al.}{2022}]{hankla2022}
{Hankla} A.~M.,  {Scepi} N.,   {Dexter} J.,  2022, \mn@doi [\mnras] {10.1093/mnras/stac1785}, \href {https://ui.adsabs.harvard.edu/abs/2022MNRAS.515..775H} {515, 775}

\bibitem[\protect\citeauthoryear{{Hogg} \& {Reynolds}}{{Hogg} \& {Reynolds}}{2017}]{hogg2017}
{Hogg} J.~D.,  {Reynolds} C.~S.,  2017, \mn@doi [\apj] {10.3847/1538-4357/aa774b}, \href {https://ui.adsabs.harvard.edu/abs/2017ApJ...843...80H} {843, 80}

\bibitem[\protect\citeauthoryear{{Hogg} \& {Reynolds}}{{Hogg} \& {Reynolds}}{2018}]{hogg2018}
{Hogg} J.~D.,  {Reynolds} C.~S.,  2018, \mn@doi [\apj] {10.3847/1538-4357/aac439}, \href {https://ui.adsabs.harvard.edu/abs/2018ApJ...861...24H} {861, 24}

\bibitem[\protect\citeauthoryear{{Kinch}, {Noble}, {Schnittman}  \& {Krolik}}{{Kinch} et~al.}{2020}]{kinch2020}
{Kinch} B.~E.,  {Noble} S.~C.,  {Schnittman} J.~D.,   {Krolik} J.~H.,  2020, \mn@doi [\apj] {10.3847/1538-4357/abc176}, \href {https://ui.adsabs.harvard.edu/abs/2020ApJ...904..117K} {904, 117}

\bibitem[\protect\citeauthoryear{{Laha}, {Ricci}, {Mather}, {Behar}, {Gallo}, {Marin}, {Mbarek}  \& {Hankla}}{{Laha} et~al.}{2025}]{laha2025}
{Laha} S.,  {Ricci} C.,  {Mather} J.~C.,  {Behar} E.,  {Gallo} L.,  {Marin} F.,  {Mbarek} R.,   {Hankla} A.,  2025, \mn@doi [Frontiers in Astronomy and Space Sciences] {10.3389/fspas.2024.1530392}, \href {https://ui.adsabs.harvard.edu/abs/2025FrASS..1130392L} {11, 1530392}

\bibitem[\protect\citeauthoryear{{Liska}, {Musoke}, {Tchekhovskoy}, {Porth}  \& {Beloborodov}}{{Liska} et~al.}{2022}]{liska2022}
{Liska} M.~T.~P.,  {Musoke} G.,  {Tchekhovskoy} A.,  {Porth} O.,   {Beloborodov} A.~M.,  2022, \mn@doi [\apjl] {10.3847/2041-8213/ac84db}, \href {https://ui.adsabs.harvard.edu/abs/2022ApJ...935L...1L} {935, L1}

\bibitem[\protect\citeauthoryear{{McConnell} et~al.,}{{McConnell} et~al.}{2002}]{mcconnell2002}
{McConnell} M.~L.,  et~al., 2002, \mn@doi [\apj] {10.1086/340436}, \href {https://ui.adsabs.harvard.edu/abs/2002ApJ...572..984M} {572, 984}

\bibitem[\protect\citeauthoryear{{Miller}, {Ryan}  \& {Dolence}}{{Miller} et~al.}{2019}]{miller2019}
{Miller} J.~M.,  {Ryan} B.~R.,   {Dolence} J.~C.,  2019, \mn@doi [\apjs] {10.3847/1538-4365/ab09fc}, \href {https://ui.adsabs.harvard.edu/abs/2019ApJS..241...30M} {241, 30}

\bibitem[\protect\citeauthoryear{{Mishra}, {Fragile}, {Johnson}  \& {Klu{\'z}niak}}{{Mishra} et~al.}{2016}]{mishra2016}
{Mishra} B.,  {Fragile} P.~C.,  {Johnson} L.~C.,   {Klu{\'z}niak} W.,  2016, \mn@doi [\mnras] {10.1093/mnras/stw2245}, \href {https://ui.adsabs.harvard.edu/abs/2016MNRAS.463.3437M} {463, 3437}

\bibitem[\protect\citeauthoryear{{Mishra}, {Fragile}, {Anderson}, {Blankenship}, {Li}  \& {Nalewajko}}{{Mishra} et~al.}{2022}]{mishra2022}
{Mishra} B.,  {Fragile} P.~C.,  {Anderson} J.,  {Blankenship} A.,  {Li} H.,   {Nalewajko} K.,  2022, \mn@doi [\apj] {10.3847/1538-4357/ac938b}, \href {https://ui.adsabs.harvard.edu/abs/2022ApJ...939...31M} {939, 31}

\bibitem[\protect\citeauthoryear{{Morales Teixeira}, {Avara}  \& {McKinney}}{{Morales Teixeira} et~al.}{2018}]{morales2018}
{Morales Teixeira} D.,  {Avara} M.~J.,   {McKinney} J.~C.,  2018, \mn@doi [\mnras] {10.1093/mnras/sty2044}, \href {https://ui.adsabs.harvard.edu/abs/2018MNRAS.480.3547M} {480, 3547}

\bibitem[\protect\citeauthoryear{{Mo{\'s}cibrodzka}, {Gammie}, {Dolence}, {Shiokawa}  \& {Leung}}{{Mo{\'s}cibrodzka} et~al.}{2009}]{moscibrodzka2009}
{Mo{\'s}cibrodzka} M.,  {Gammie} C.~F.,  {Dolence} J.~C.,  {Shiokawa} H.,   {Leung} P.~K.,  2009, \mn@doi [\apj] {10.1088/0004-637X/706/1/497}, \href {https://ui.adsabs.harvard.edu/abs/2009ApJ...706..497M} {706, 497}

\bibitem[\protect\citeauthoryear{{Mummery}, {Ingram}, {Davis}  \& {Fabian}}{{Mummery} et~al.}{2024}]{mummery2024}
{Mummery} A.,  {Ingram} A.,  {Davis} S.,   {Fabian} A.,  2024, \mn@doi [\mnras] {10.1093/mnras/stae1160}, \href {https://ui.adsabs.harvard.edu/abs/2024MNRAS.531..366M} {531, 366}

\bibitem[\protect\citeauthoryear{{Naethe Motta}, {Jacquemin-Ide}, {Nemmen}, {Liska}  \& {Tchekhovskoy}}{{Naethe Motta} et~al.}{2025}]{motta2025}
{Naethe Motta} P.,  {Jacquemin-Ide} J.,  {Nemmen} R.,  {Liska} M. T.~P.,   {Tchekhovskoy} A.,  2025, \mn@doi [arXiv e-prints] {10.48550/arXiv.2505.08855}, \href {https://ui.adsabs.harvard.edu/abs/2025arXiv250508855N} {p. arXiv:2505.08855}

\bibitem[\protect\citeauthoryear{{Narayan} \& {Yi}}{{Narayan} \& {Yi}}{1994}]{narayan1994}
{Narayan} R.,  {Yi} I.,  1994, \mn@doi [\apjl] {10.1086/187381}, \href {https://ui.adsabs.harvard.edu/abs/1994ApJ...428L..13N} {428, L13}

\bibitem[\protect\citeauthoryear{{Noble}, {Gammie}, {McKinney}  \& {Del Zanna}}{{Noble} et~al.}{2006}]{noble2006}
{Noble} S.~C.,  {Gammie} C.~F.,  {McKinney} J.~C.,   {Del Zanna} L.,  2006, \mn@doi [\apj] {10.1086/500349}, \href {https://ui.adsabs.harvard.edu/abs/2006ApJ...641..626N} {641, 626}

\bibitem[\protect\citeauthoryear{{Noble}, {Krolik}  \& {Hawley}}{{Noble} et~al.}{2009}]{noble2009}
{Noble} S.~C.,  {Krolik} J.~H.,   {Hawley} J.~F.,  2009, \mn@doi [\apj] {10.1088/0004-637X/692/1/411}, \href {https://ui.adsabs.harvard.edu/abs/2009ApJ...692..411N} {692, 411}

\bibitem[\protect\citeauthoryear{{Novikov} \& {Thorne}}{{Novikov} \& {Thorne}}{1973}]{novikovthorne}
{Novikov} I.~D.,  {Thorne} K.~S.,  1973, in {Dewitt} C.,  {Dewitt} B.~S.,  eds, Black Holes (Les Astres Occlus). pp 343--450

\bibitem[\protect\citeauthoryear{{Page} \& {Thorne}}{{Page} \& {Thorne}}{1974}]{page1974}
{Page} D.~N.,  {Thorne} K.~S.,  1974, \mn@doi [\apj] {10.1086/152990}, \href {https://ui.adsabs.harvard.edu/abs/1974ApJ...191..499P} {191, 499}

\bibitem[\protect\citeauthoryear{{Penna}, {McKinney}, {Narayan}, {Tchekhovskoy}, {Shafee}  \& {McClintock}}{{Penna} et~al.}{2010}]{penna2010}
{Penna} R.~F.,  {McKinney} J.~C.,  {Narayan} R.,  {Tchekhovskoy} A.,  {Shafee} R.,   {McClintock} J.~E.,  2010, \mn@doi [\mnras] {10.1111/j.1365-2966.2010.17170.x}, \href {https://ui.adsabs.harvard.edu/abs/2010MNRAS.408..752P} {408, 752}

\bibitem[\protect\citeauthoryear{{Ressler}, {Tchekhovskoy}, {Quataert}, {Chandra}  \& {Gammie}}{{Ressler} et~al.}{2015}]{ressler2015}
{Ressler} S.~M.,  {Tchekhovskoy} A.,  {Quataert} E.,  {Chandra} M.,   {Gammie} C.~F.,  2015, \mn@doi [\mnras] {10.1093/mnras/stv2084}, \href {https://ui.adsabs.harvard.edu/abs/2015MNRAS.454.1848R} {454, 1848}

\bibitem[\protect\citeauthoryear{{Ressler}, {Tchekhovskoy}, {Quataert}  \& {Gammie}}{{Ressler} et~al.}{2017}]{ressler2017}
{Ressler} S.~M.,  {Tchekhovskoy} A.,  {Quataert} E.,   {Gammie} C.~F.,  2017, \mn@doi [\mnras] {10.1093/mnras/stx364}, \href {https://ui.adsabs.harvard.edu/abs/2017MNRAS.467.3604R} {467, 3604}

\bibitem[\protect\citeauthoryear{{Reynolds}}{{Reynolds}}{2021}]{reynolds2021}
{Reynolds} C.~S.,  2021, \mn@doi [\araa] {10.1146/annurev-astro-112420-035022}, \href {https://ui.adsabs.harvard.edu/abs/2021ARA&A..59..117R} {59, 117}

\bibitem[\protect\citeauthoryear{{Ryan}, {Dolence}  \& {Gammie}}{{Ryan} et~al.}{2015}]{ryan2015}
{Ryan} B.~R.,  {Dolence} J.~C.,   {Gammie} C.~F.,  2015, \mn@doi [\apj] {10.1088/0004-637X/807/1/31}, \href {https://ui.adsabs.harvard.edu/abs/2015ApJ...807...31R} {807, 31}

\bibitem[\protect\citeauthoryear{{Ryan}, {Ressler}, {Dolence}, {Tchekhovskoy}, {Gammie}  \& {Quataert}}{{Ryan} et~al.}{2017}]{ryan2017}
{Ryan} B.~R.,  {Ressler} S.~M.,  {Dolence} J.~C.,  {Tchekhovskoy} A.,  {Gammie} C.,   {Quataert} E.,  2017, \mn@doi [\apjl] {10.3847/2041-8213/aa8034}, \href {https://ui.adsabs.harvard.edu/abs/2017ApJ...844L..24R} {844, L24}

\bibitem[\protect\citeauthoryear{{Scepi}, {Begelman}  \& {Dexter}}{{Scepi} et~al.}{2024a}]{scepi2024}
{Scepi} N.,  {Begelman} M.~C.,   {Dexter} J.,  2024a, \mn@doi [\mnras] {10.1093/mnras/stad3299}, \href {https://ui.adsabs.harvard.edu/abs/2024MNRAS.527.1424S} {527, 1424}

\bibitem[\protect\citeauthoryear{{Scepi}, {Dexter}, {Begelman}, {Marcel}, {Ferreira}  \& {Petrucci}}{{Scepi} et~al.}{2024b}]{scepi2024b}
{Scepi} N.,  {Dexter} J.,  {Begelman} M.~C.,  {Marcel} G.,  {Ferreira} J.,   {Petrucci} P.-O.,  2024b, \mn@doi [\aap] {10.1051/0004-6361/202451568}, \href {https://ui.adsabs.harvard.edu/abs/2024A&A...692A.153S} {692, A153}

\bibitem[\protect\citeauthoryear{{Shafee}, {McKinney}, {Narayan}, {Tchekhovskoy}, {Gammie}  \& {McClintock}}{{Shafee} et~al.}{2008}]{shafee2008}
{Shafee} R.,  {McKinney} J.~C.,  {Narayan} R.,  {Tchekhovskoy} A.,  {Gammie} C.~F.,   {McClintock} J.~E.,  2008, \mn@doi [\apjl] {10.1086/593148}, \href {https://ui.adsabs.harvard.edu/abs/2008ApJ...687L..25S} {687, L25}

\bibitem[\protect\citeauthoryear{{Shakura} \& {Sunyaev}}{{Shakura} \& {Sunyaev}}{1973}]{ss73}
{Shakura} N.~I.,  {Sunyaev} R.~A.,  1973, \aap, \href {https://ui.adsabs.harvard.edu/abs/1973A&A....24..337S} {24, 337}

\bibitem[\protect\citeauthoryear{{Stepney} \& {Guilbert}}{{Stepney} \& {Guilbert}}{1983}]{stepney1983}
{Stepney} S.,  {Guilbert} P.~W.,  1983, \mn@doi [\mnras] {10.1093/mnras/204.4.1269}, \href {https://ui.adsabs.harvard.edu/abs/1983MNRAS.204.1269S} {204, 1269}

\bibitem[\protect\citeauthoryear{{Uttley}, {Cackett}, {Fabian}, {Kara}  \& {Wilkins}}{{Uttley} et~al.}{2014}]{uttley2014}
{Uttley} P.,  {Cackett} E.~M.,  {Fabian} A.~C.,  {Kara} E.,   {Wilkins} D.~R.,  2014, \mn@doi [\aapr] {10.1007/s00159-014-0072-0}, \href {https://ui.adsabs.harvard.edu/abs/2014A&ARv..22...72U} {22, 72}

\bibitem[\protect\citeauthoryear{{Weisskopf} et~al.,}{{Weisskopf} et~al.}{2016}]{ixpe}
{Weisskopf} M.~C.,  et~al., 2016, in {den Herder} J.-W.~A.,  {Takahashi} T.,   {Bautz} M.,  eds,  Society of Photo-Optical Instrumentation Engineers (SPIE) Conference Series Vol. 9905, Space Telescopes and Instrumentation 2016: Ultraviolet to Gamma Ray. p. 990517, \mn@doi{10.1117/12.2235240}

\bibitem[\protect\citeauthoryear{{Werner}, {Uzdensky}, {Begelman}, {Cerutti}  \& {Nalewajko}}{{Werner} et~al.}{2018}]{werner2018}
{Werner} G.~R.,  {Uzdensky} D.~A.,  {Begelman} M.~C.,  {Cerutti} B.,   {Nalewajko} K.,  2018, \mn@doi [\mnras] {10.1093/mnras/stx2530}, \href {https://ui.adsabs.harvard.edu/abs/2018MNRAS.473.4840W} {473, 4840}

\bibitem[\protect\citeauthoryear{{Yuan} \& {Narayan}}{{Yuan} \& {Narayan}}{2014}]{yuan2014}
{Yuan} F.,  {Narayan} R.,  2014, \mn@doi [\araa] {10.1146/annurev-astro-082812-141003}, \href {https://ui.adsabs.harvard.edu/abs/2014ARA&A..52..529Y} {52, 529}

\bibitem[\protect\citeauthoryear{{Zhu}, {Davis}, {Narayan}, {Kulkarni}, {Penna}  \& {McClintock}}{{Zhu} et~al.}{2012}]{zhu2012}
{Zhu} Y.,  {Davis} S.~W.,  {Narayan} R.,  {Kulkarni} A.~K.,  {Penna} R.~F.,   {McClintock} J.~E.,  2012, \mn@doi [\mnras] {10.1111/j.1365-2966.2012.21181.x}, \href {https://ui.adsabs.harvard.edu/abs/2012MNRAS.424.2504Z} {424, 2504}

\bibitem[\protect\citeauthoryear{{Zoghbi} \& {Fabian}}{{Zoghbi} \& {Fabian}}{2011}]{zoghbi2011}
{Zoghbi} A.,  {Fabian} A.~C.,  2011, \mn@doi [\mnras] {10.1111/j.1365-2966.2011.19655.x}, \href {https://ui.adsabs.harvard.edu/abs/2011MNRAS.418.2642Z} {418, 2642}

\makeatother
\end{thebibliography}

\appendix

\section{Two-temperature thermalization tests} \label{app:tt-tests}
\begin{figure*}
    \centering
    \includegraphics[width=\textwidth]{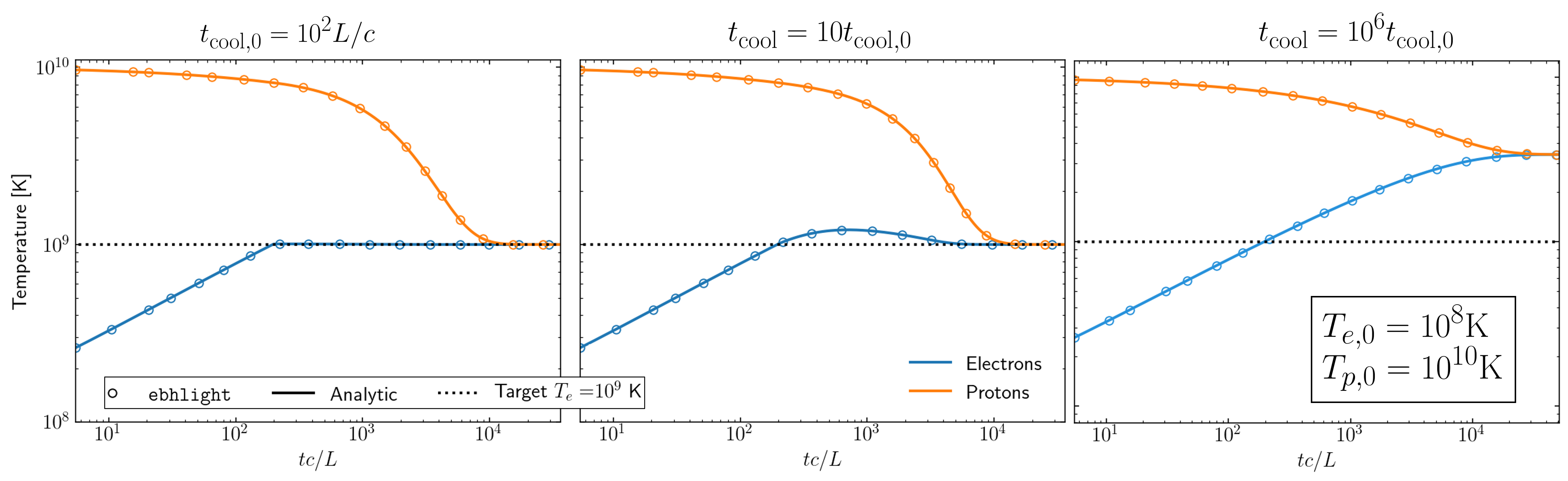}
    \caption{Tests of the electron-only cooling function agree well with analytic expectations. }
    \label{fig:gasbox}
\end{figure*}
To test the implementation of the electron-only cooling function and the implicit Coulomb solver, we simulate a 2D, doubly-periodic gas box where the protons and electrons have initially different temperatures. 
Such a set-up eliminates viscous heating and allows us to explore how Coulomb coupling and electron cooling interact. 
These test simulations have a constant $t_{\rm cool}$, a uniform density, no magnetic field, and a target temperature of $10^9$ K.

We compare our~$\texttt{nubhlight}$ results to the solution obtained from the descriptive equations: 
\begin{align}
    \frac{dT_e}{dt}&=\frac{(\gamma_e-1)}{n_0k_B}\left(Q_{\rm coul}-Q_{\rm cool}\right)\\
    \frac{dT_p}{dt}&=-q_C\frac{(\gamma_p-1)}{n_0k_B}.
\end{align}
In the above, $Q_{\rm coul}$ is the rate of energy exchange between electrons and protons per unit volume~\citep{stepney1983}. 

We show the results from three typical simulations in Figure~\ref{fig:gasbox}. 
All three simulations start with an electron temperature $T_{e,0}=10^8$ K and a proton temperature $T_{p,0}=10^{10}$ K. 
In the first simulation, $n_0=6\times10^{21} {\rm particles/cm^3}$ and $t_{\rm cool}=10^2~L/c$. 
With these parameters, the electrons reach the target cooling temperature before the electrons and protons thermalize. 
Because the cooling time is short, the electrons stay glued to the target temperature. 
The second simulation has a cooling time ten times that of the first, allowing the electron temperature to temporarily overshoot the target temperature. 
In the last simulation, a density ten times that of the first two simulations means that Coulomb collisions are much stronger initially.
Coulomb collisions therefore force the gas and electron fluid to thermalize much faster. 
Because of the stronger Coulomb collisions, the electron temperature briefly exceeds the target temperature despite a shorter cooling time than the second simulation. 
See~\citet{motta2025} for a detailed comparison of numerical techniques.

%\section{Enhanced Coulomb collisions}
\section{Collapse to a Single-Temperature Steady State at High Accretion Rate} \label{app:enhancedQ}
\begin{figure*}
    \centering
    \includegraphics[width=0.95\linewidth]{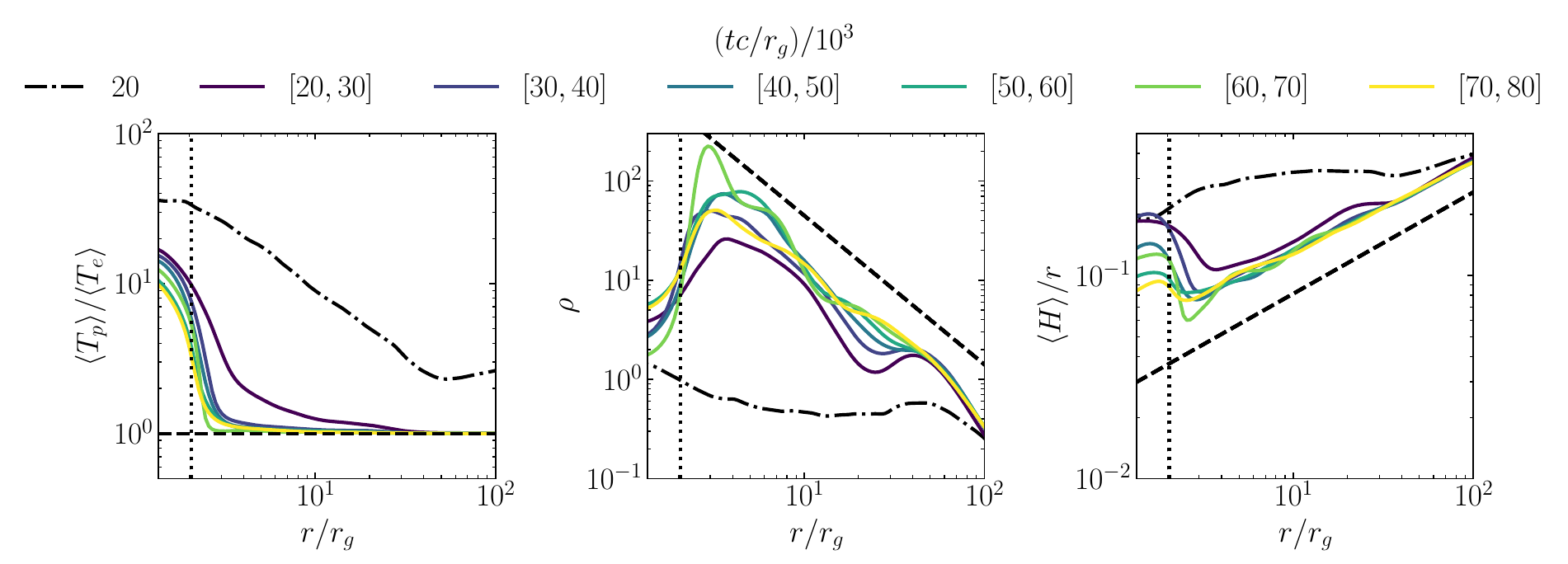}
    \caption{Density-weighted shell averages showing the collapse of a hot, two-temperature accretion flow to a thin, single-temperature accretion disc. The black dash-dot line shows the initial condition of a two-temperature torus. Colored lines show the evolution in $10^4r_g/c$ time-averaged increments. Left panel: proton-to-electron temperature ratio radial profile, with the dashed black line showing the single-temperature regime with $T_p=T_e$. Center panel: mass density radial profile, with the dashed black line showing $\rho\sim r^{-3/2}$. Right panel: disc scale height ratio radial profile, with the dashed black line showing the theoretical scaling of $H/r\sim r^{1/2}$ for a constant target temperature.}
    \label{fig:thin-grid}
\end{figure*}
Much of the collapse occurs over the course of the first $10^4r_g/c$ after restart from the low accretion rate. 
Figure~\ref{fig:thin-grid} shows the evolution over time of a low-resolution M2 simulation, time-averaged over $10^4r_g/c$ increments.
The initial profiles from the M6 simulation at $2\times10^4r_g/c$, shown as black dash-dot lines in each panel, demonstrate how dramatically the disc structure changes within the first $10^4r_g/c$ (purple line). 
Whereas the hot accretion flow clearly demonstrates two-temperature structure in the ratio $T_p/T_e>2$ at all radii shown, the higher accretion rate M2 quickly develops into a single-temperature regime for $r>r_{\rm ISCO}$ (Fig.~\ref{fig:thin-grid} left panel).
Similarly, the mass density radial profile changes from roughly constant with radius to sharply decreasing with radius; see Fig.~\ref{fig:thin-grid} center panel.
The density scale height radial profile also reflects the change in disc structure.
Initially, $H/r$ remained approximately 0.3 for the radii shown. 
After the restart with higher accretion rate, the density scale height dropped to $H/r<0.1$ and follows the radial scaling $H/r\sim r^{1/2}$ expected for a thin disc with a constant target temperature (black dashed line), shown in Fig.~\ref{fig:thin-grid} right panel.

%\subsection{Justification for Enhanced Coulomb Collisions}

%\subsection{Steady State?}
% ------------------------------------------------------- %

%%%%%%%%%%%%%%%%%%%%%%%%%%%%%%%%%%%%%%%%%%%%%%%%%%
% Don't change these lines
\bsp	% typesetting comment
\label{lastpage}
\end{document}